\documentclass[aps,prd,onecolumn,notitlepage,nofootinbib,showpacs,showkeys,preprintnumbers,superscriptaddress,groupedaddress]{revtex4-2}
\usepackage[T1]{fontenc}
\usepackage[utf8]{inputenc}
\usepackage{bm}
\usepackage{mathtools}
\usepackage{esint}
\usepackage{xcolor}
\usepackage{array}
\usepackage{float}
\usepackage{graphicx}
\PassOptionsToPackage{normalem}{ulem}
\usepackage{ulem}
\usepackage{hyperref}
\usepackage{multirow}

\begin{document}
	
	\title{Spatially covariant gravity with two degrees of freedom:\\ perturbative analysis}

	\author{Yu-Min Hu}%
	\email[Email: ]{huym25@mail2.sysu.edu.cn}
	\affiliation{%
	School of Physics and Astronomy, Sun Yat-sen University, Guangzhou 510275, China}

	\author{Xian Gao}%
	\email[Corresponding author: ]{gaoxian@mail.sysu.edu.cn}
	\affiliation{%
	School of Physics and Astronomy, Sun Yat-sen University, Guangzhou 510275, China}

	\date{April 16, 2021}
	
	\begin{abstract}
		We revisit the problem of building the Lagrangian of a large class of metric theories that respect spatial covariance, which propagate at most two degrees of freedom and in particular no scalar mode. The Lagrangians are polynomials built of the spatially covariant geometric quantities. By expanding the Lagrangian around a cosmological background and focusing on the scalar modes only, we find the conditions for the coefficients of the monomials in order to eliminate the scalar mode at the linear order in perturbations.
		We find the conditions up to $d=4$ with $d$ the total number of derivatives in the monomials and determine the explicit Lagrangians for the cases of $d=2$, $d=3$ as well as the combination of $d=2$ and $d=3$. We also expand the Lagrangian of $d=2$ to the cubic order in perturbations, and find additional conditions for the coefficients such that the scalar mode is eliminated up to the cubic order. This perturbative analysis can be performed order by order, and one expects to determine the final Lagrangian at some finite order such that the scalar mode is fully eliminated. Our analysis provides an alternative and complimentary approach to building spatially covariant gravity with only tensorial degrees of freedom.
		The resulting theories can be used as alternatives to the general relativity to describe the tensorial gravitational waves in a cosmological setting.
	\end{abstract}

\maketitle


\section{Introduction}

Recently, there is revival of the interest of questioning the uniqueness of general relativity (GR) as the theory of two tensorial degrees of freedom (TTDOF's). 
The Lovelock's theorem \cite{Lovelock:1972vz} is an answer to this question, which claims that GR is the unique four dimensional theory for the spacetime metric with the second order equations of motion, which obeys the general covariance and locality. As a result, GR is the unique theory for the TTDOF's if all the assumptions of Lovelock are preserved.

From the field theoretical point of view, the idea of embedding the gravitational degrees of freedom in a field theory of metric variables was also explored. By coupling a massless spin-2 field to the energy–momentum tensors of matter field(s) as well as of its own, it was arguably believed that the Einstein-Hilbert action is the unique theory one will arrive at.
This approach can be traced back to the Fierz-Pauli theory \cite{Fierz:1939ix} and was further widely developed in \cite{Gupta:1954zz,Kraichnan:1955zz,Kraichnan:1956pug,Gupta:1957etg,Feynman:1996kb,Weinberg:1964kqu,Weinberg:1964ew,Deser:1969wk,Teitelboim:1972vw,Boulware:1974sr,Wald:1986bj,Babak:1999dc,Barbour:2000qg,Padmanabhan:2004xk,Deser:2009fq} (see \cite{Alvarez:1988tb} for a review)\footnote{Recently the ``bootstrap'' approach also provides new understandings on the uniqueness of GR \cite{Britto:2005fq,Benincasa:2007qj,ArkaniHamed:2008yf,Krasnov:2014eza,Carballo-Rubio:2018bmu,Pajer:2020wnj} (see \cite{Benincasa:2013faa} for a review).}.

In this work, we shall examining the conditions of propagating only TTDOF's in a large class of metric theories respecting spatial covariance, which we dub the spatially covariant gravity (SCG).
The SCG can be traced back to the ghost condensation \cite{ArkaniHamed:2003uy} and was developed in the effective field theory of inflation \cite{Cheung:2007st,Gubitosi:2012hu} as well as in the Ho\v{r}ava gravity \cite{Horava:2009uw,Horava:2008ih}.
It was further generalized in \cite{Gao:2014soa} in which a large class of SCG theories was proposed and was extended in \cite{Gao:2018znj} by including the dynamical lapse function and in \cite{Gao:2018izs} with an auxiliary scalar field.

Theories different from GR while propagating only TTDOF's firstly arose in the so-called ``cuscuton'' theory \cite{Afshordi:2006ad,Afshordi:2007yx} and in a sub-class of Ho\v{r}ava gravity \cite{Zhu:2011xe,Zhu:2011yu}.
In \cite{Lin:2017oow} a class of SCG theories with only TTDOF's was proposed as the minimal modification of the GR (MMG).
The cuscuton and MMG theories have been further extended \cite{Chagoya:2018yna,Iyonaga:2018vnu,Lin:2018mip,Mukohyama:2019unx,Lin:2019ijq,DeFelice:2020eju,Aoki:2021zuy} and their applications on cosmology and black holes  have been widely studied \cite{Carballo-Rubio:2018czn,Ito:2019ztb,Quintin:2019orx,Iyonaga:2020bmm,DeFelice:2020cpt,DeFelice:2020onz,DeFelice:2020prd,Hagala:2020eax,Sangtawee:2021mhz}.
A class of 4-dimension Einstein-Gauss-Bonnet gravity was also proposed recently as an arguable TTDOF  theory \cite{Glavan:2019inb}.

We shall employ the framework of SCG due to the following reasons.
\begin{itemize}
	\item The Lagrangians of SCG theories are automatically written in the spacetime-split form, which are convenient for analysing the time evolution and the degrees of freedom using either the equations of motion or Hamiltonian constraint analysis.
	
	\item The SCG can be viewed as the gauge fixed version of the scalar-tensor theory with a single timelike scalar field. By choosing the time coordinate as the scalar field $t=t(\phi)$, which is dubbed the unitary gauge in the literature, the generally covariant single field scalar-tensor theory can be naturally recast in the form of a SCG theory. Therefore the SCG can be used as a ``generator'' of the scalar-tensor theory, especially when the higher derivatives are present. 
	The generally covariant scalar-tensor monomials and SCG monomials have been classified and their correspondence has been investigated in
	\cite{Gao:2020juc,Gao:2020yzr,Gao:2020qxy}\footnote{The correspondence is subtle when the unitary gauge is not accessible, see \cite{DeFelice:2018ewo} for a discussion.}.
	This may help to build well-behaved higher derivative scalar-tensor theories after the (re)construction of the theory of Horndeski \cite{Horndeski:1974wa} in its modern form \cite{Nicolis:2008in,Deffayet:2011gz,Kobayashi:2011nu} and the degenerate higher order derivative scalar-tensor theory \cite{Gleyzes:2014dya,Gleyzes:2014qga,Langlois:2015cwa,Motohashi:2016ftl} (see \cite{Langlois:2018dxi,Kobayashi:2019hrl} for reviews and references therein).
	
	\item Thanks to the spacetime-splitting nature of the SCG, the construction of SCG with only a single scalar degree of freedom becomes virtually trivial. Indeed, as being shown in \cite{Lin:2014jga,Gao:2014fra,Saitou:2016lvb}, the SCG without the dynamical lapse function automatically propagates at most three degrees of freedom.
	When the lapse function becomes dynamical, more conditions must be imposed \cite{Gao:2018znj}.
\end{itemize}

Counting the number of degrees of freedom can be well-performed through Dirac's Hamiltonian constraint analysis (see \cite{Henneaux:1992ig} for a comprehensive review).
In the framework of SCG, there are in principle two approaches to finding the conditions of propagating at most two degrees of freedom. 
\begin{itemize}
	\item The traditional and conservative approach, is to start from the Lagrangian and perform the Legendre transformation to derive the Hamiltonian, then to find the conditions for the Lagrangian by performing the Hamiltonian constraint analysis.
	In \cite{Gao:2019twq}, starting from a general local Lagrangian of SCG, conditions of propagating at most two degrees of freedom have been derived. 
	The analysis was also generalized in \cite{Lin:2020nro} with a dynamical lapse function.
	
	\item The other approach is to start with the Hamiltonian directly, and to determine the conditions for the Hamiltonian instead of the Lagrangian.
	Indeed, simplified structure and condition(s) are found at the level of Hamiltonian in \cite{Mukohyama:2019unx} in a class of SCG theories.
	This approach can be made even more ``trivial'' by imposing additional constraint(s) in the phase space through auxiliary variable(s) \cite{Yao:2020tur}.
\end{itemize}
Both approaches have their merits and shortcomings.
For the ``Lagrangian'' approach, it is more convenient to deal with the \emph{local} Lagrangian, while the conditions are functional differential equations for the Lagrangian and mathematically complicated to be solved.
For the ``Hamiltonian'' approach, one is able to determine the Hamiltonian in a simple manner, while the corresponding Lagrangian is involved and typically \emph{non-local} due to the presence of extra auxiliary variables.

In view of the above considerations, in this work we employ an alternative approach to constructing the SCG with only TTDOF's.
We shall deal with the Lagrangian directly and determine the conditions at the level of equations of motion.  
In fact, constraint analysis as well as counting the number of degrees of freedom can be equivalently performed at the level of the Lagrangian and the equations of motion \cite{Mukunda:1974dr,Rothe:2010dzf}.

The idea is based on the fact that if the Lagrangian propagates at most two DOF's --- and in particular, no scalar mode --- at the non-perturbative level, the scalar mode must not show up at any finite order in the perturbative expansion around a spatially homogeneous and isotropic background.
In particular, the conditions can be determined order by order in a perturbative analysis, which may be relatively easier to be managed.
This is also the approach in \cite{Iyonaga:2018vnu} to building the so-called ``extended cuscuton'' theory.
The same idea was also employed in \cite{Gao:2019lpz} to find conditions for the SCG Lagrangians quadratic in the extrinsic curvature and in  the velocity of the lapse function to propagate at most three degrees of freedom.

This work is organized as follows.
In Sec. \ref{sec:scg} we briefly review the spatially covariant gravity and the general conditions to have at most two degrees of freedom.
In Sec. \ref{sec:pert_degen} we describe our perturbative approach and derive the degeneracy condition, in order to eliminate the scalar mode at linear order in perturbations.
In Sec. \ref{sec:quad} we use the degeneracy condition to find the conditions for the Lagrangians up to $d=4$ and give the explicit Lagrangians for $d=2$, $d=3$ as well as the combination of $d=2,3$.
In Sec. \ref{sec:cub} we use the Lagrangian of $d=2$ as an illustrative example to show how to eliminate the scalar mode at the next order in perturbations.
We summarize our results in Sec. \ref{sec:con}.

\section{Spatially covariant gravity with 2 degrees of freedom} \label{sec:scg}

In this section, we make a brief review of the framework of spatially covariant gravity (SCG) theory, and in particular the classification of the SCG polynomials.
We also briefly summarize the conditions of having only two degrees of freedom, which are derived in \cite{Gao:2019twq}.

The action of the spatially covariant gravity theories takes the general form
\begin{equation} 
	S=\int \mathrm{d}t\mathrm{d}^{3}x\,N\sqrt{h}\mathcal{L}\left(t,N,h_{ij},K_{ij},R_{ij},\nabla_{i},\varepsilon_{ijk}\right), \label{S_form}
\end{equation}
where $N$ is the lapse function, $h_{ij}$ is the 3-dimensional spatial metric, $K_{ij}$ is the extrinsic curvature defined by
	\begin{equation} 
		K_{ij}=\frac{1}{2N}\left(\partial_{t}{h}_{ij}-\pounds_{\!\vec{N}}h_{ij}\right),
	\end{equation}
with $\pounds_{\!\vec{N}}$ the Lie derivative with respect to the shift vector $N^{i}$, $R_{ij}$ is the 3-dimensional spatial Ricci tensor, $\nabla_{i}$ is the covariant derivative compatible with the spatial metric $h_{ij}$.  
The spatial Levi-Civita tensor $\varepsilon_{ijk}=\sqrt{h}\epsilon_{ijk}$ with $\epsilon_{123}=1$ is allowed, thus the parity-violating terms can be constructed by those building blocks with $\varepsilon_{ijk}$\footnote{The spatially covariant parity violating terms and their cosmological implications have been widely investigated, see (e.g.) \cite{Qiao:2019wsh,Zhao:2019xmm,Zhao:2019szi,Qiao:2019hkz,Wang:2020cub,Li:2020xjt,Li:2021wij,Gao:2019liu}.}. 
Note in principle the lapse function $N$ may also acquire a kinetic term through $\frac{1}{N}\left(\partial_{t}{N}-N^{i}\nabla_{i}N\right)$, which has been considered in \cite{Gao:2018znj}.
The shift vector $N_{i}$ by itself is not a genuine geometric quantity of the spacetime foliation structure, which merely encodes the gauge freedom of choosing the spatial coordinates.

In this work, instead of analysing a general Lagrangian as in (\ref{S_form}), we concentrat on polynomial-type Lagrangians, which are linear combinations of the SCG monomials. 
The irreducible SCG monomials are exhausted and classified up to $d=4$ in \cite{Gao:2020yzr} with $d$ the total number of derivatives in the monomials. Here we briefly describe the construction with improved notations following \cite{Gao:2020qxy}.
We assign each SCG model by a set of integers $(c_{0};d_{2},d_{3})$ according to their corresponding monomials in the scalar-tensor theories after the Stueckelberg mapping. Precisely, $c_{0}$ is the number of spacetime Riemann curvature tensor, $d_{2}$, $d_{3}$ are numbers of the second, the third order generally covariant derivatives of the scalar field $\phi$, respectively. 
In fact we have the simple correspondences
	\begin{eqnarray} 
		K_{ij}\sim a_{i}	& \sim &(0;1,0),
		\\
		R_{ij}	& \sim & \left(1;0,0\right),
		\\
		\nabla_{k}K_{ij}\sim\nabla_{i}a_{j}	&\sim & \left(0;0,1\right),
	\end{eqnarray}
and thus $d$ can be expressed by
	\begin{equation} 
		d=\underset{n=0}{\sum}\left[\left(n+2\right)c_{n}+\left(n+1\right)d_{n+2}\right].
	\end{equation}
We thus classify the various SCG monomials with $d$ and then with the categories labelled by $(c_{0};d_{2},d_{3})$.

Up to $d=4$, the Lagrangian built of the irreducible SCG monomials is
	\begin{equation} 
	\mathcal{L}=\mathcal{L}^{\left(0\right)}+\mathcal{L}^{\left(1\right)}+\mathcal{L}^{\left(2\right)}+\mathcal{L}^{\left(3\right)}+\mathcal{L}^{\left(4\right)}+\tilde{\mathcal{L}}^{\left(3\right)}+\tilde{\mathcal{L}}^{\left(4\right)}, \label{Lag_SCG}
	\end{equation}
where the parity-preserving terms are
\begin{equation}
	\mathcal{L}^{\left(0\right)}=c_{1}^{\left(0;0,0\right)},
\end{equation}
\begin{equation}
	\mathcal{L}^{\left(1\right)}=c_{1}^{\left(0;1,0\right)}K,
\end{equation}
\begin{equation}
	\mathcal{L}^{\left(2\right)}=c_{1}^{\left(0;2,0\right)}K_{ij}K^{ij}+c_{2}^{\left(0;2,0\right)}a_{i}a^{i}+c_{3}^{\left(0;2,0\right)}K^{2}+c_{1}^{\left(1;0,0\right)}R, \label{calL_2}
\end{equation}
\begin{eqnarray}
	\mathcal{L}^{\left(3\right)} & = & c_{1}^{\left(0;3,0\right)}K_{ij}K^{jk}K_{k}^{i}+c_{2}^{\left(0;3,0\right)}K_{ij}a^{i}a^{j}+c_{3}^{\left(0;3,0\right)}K_{ij}K^{ij}K+c_{4}^{\left(0;3,0\right)}Ka_{i}a^{i}+c_{5}^{\left(0;3,0\right)}K^{3}\nonumber \\
	&  & +c_{1}^{\left(0;1,1\right)}K_{ij}\nabla^{i}a^{j}+c_{2}^{\left(0;1,1\right)}K\nabla_{i}a^{i}\nonumber \\
	&  & +c_{1}^{\left(1;1,0\right)}R^{ij}K_{ij}+c_{2}^{\left(1;1,0\right)}RK, \label{calL_3}
\end{eqnarray}
\begin{eqnarray}
	\mathcal{L}^{\left(4\right)} & = & c_{1}^{\left(0;4,0\right)}K_{ik}K_{j}^{k}a^{i}a^{j}+c_{2}^{\left(0;4,0\right)}K_{ij}K^{jk}K_{k}^{i}K+c_{3}^{\left(0;4,0\right)}K_{ij}a^{i}a^{j}K+c_{4}^{\left(0;4,0\right)}\left(K_{ij}K^{ij}\right)^{2}\nonumber \\
	&  & +c_{5}^{\left(0;4,0\right)}K_{ij}K^{ij}a_{k}a^{k}+c_{6}^{\left(0;4,0\right)}\left(a_{i}a^{i}\right)^{2}+c_{7}^{\left(0;4,0\right)}K_{ij}K^{ij}K^{2}+c_{8}^{\left(0;4,0\right)}a_{i}a^{i}K^{2}+c_{9}^{\left(0;4,0\right)}K^{4}\nonumber \\
	&  & +c_{1}^{\left(0;2,1\right)}K_{i}^{k}K_{jk}\nabla^{i}a^{j}+c_{2}^{\left(0;2,1\right)}K_{j}^{i}a^{j}\nabla_{k}K_{i}^{k}+c_{3}^{\left(0;2,1\right)}K_{j}^{i}a^{j}\nabla_{i}K+c_{4}^{\left(0;2,1\right)}K_{ij}\nabla^{i}a^{j}K\nonumber \\
	&  & +c_{5}^{\left(0;2,1\right)}K_{ij}K^{ij}\nabla_{k}a^{k}+c_{6}^{\left(0;2,1\right)}a_{i}a^{i}\nabla_{j}a^{j}+c_{7}^{\left(0;2,1\right)}K^{2}\nabla_{i}a^{i}\nonumber \\
	&  & +c_{1}^{\left(0;0,2\right)}\nabla_{k}K_{ij}\nabla^{k}K^{ij}+c_{2}^{\left(0;0,2\right)}\nabla_{i}K^{ij}\nabla_{k}K_{j}^{k}+c_{3}^{\left(0;0,2\right)}\nabla_{i}K^{ij}\nabla_{j}K\nonumber \\
	&  & +c_{4}^{\left(0;0,2\right)}\nabla_{i}K\nabla^{i}K+c_{5}^{\left(0;0,2\right)}\nabla_{i}a_{j}\nabla^{i}a^{j}+c_{6}^{\left(0;0,2\right)}\left(\nabla_{i}a^{i}\right)^{2}\nonumber \\
	&  & +c_{1}^{\left(1;2,0\right)}R_{ij}K_{k}^{i}K^{jk}+c_{2}^{\left(1;2,0\right)}R_{ij}a^{i}a^{j}+c_{3}^{\left(1;2,0\right)}R_{ij}K^{ij}K+c_{4}^{\left(1;2,0\right)}RK_{ij}K^{ij}+c_{5}^{\left(1;2,0\right)}Ra_{i}a^{i}+c_{6}^{\left(1;2,0\right)}RK^{2}\nonumber \\
	&  & +c_{1}^{\left(2;0,0\right)}R_{ij}R^{ij}+c_{2}^{\left(2;0,0\right)}R^{2}\nonumber \\
	&  & +c_{1}^{\left(1;0,1\right)}R\nabla_{i}a^{i}, \label{calL_4}
\end{eqnarray}
and the parity-violating terms are
	\begin{eqnarray}
		\tilde{\mathcal{L}}^{\left(3\right)} & =\tilde{c}_{1}^{\left(0;1,1\right)} & \varepsilon_{ijk}K_{l}^{i}\nabla^{j}K^{kl}, \label{calL_3p}
	\end{eqnarray}
	\begin{eqnarray}
		\tilde{\mathcal{L}}^{\left(4\right)} & = & \tilde{c}_{1}^{\left(0;2,1\right)}\varepsilon_{ijk}K^{im}K^{jn}\nabla_{m}K_{n}^{k}+\tilde{c}_{2}^{\left(0;2,1\right)}\varepsilon_{ijk}K^{mn}K_{m}^{i}\nabla^{j}K_{n}^{k}+\tilde{c}_{3}^{\left(0;2,1\right)}\varepsilon_{ijk}K_{l}^{i}a^{j}\nabla^{k}a^{l}+\tilde{c}_{4}^{\left(0;2,1\right)}\varepsilon_{ijk}K_{l}^{i}\nabla^{j}K^{kl}K\nonumber \\
		&  & +\tilde{c}_{1}^{\left(1;2,0\right)}\varepsilon_{ijk}R_{l}^{i}K^{jl}a^{k}+\tilde{c}_{1}^{\left(1;0,1\right)}\varepsilon_{ijk}R_{l}^{i}\nabla^{j}K^{kl}. \label{calL_4p}
	\end{eqnarray}
In the above, $a_{i}\equiv\nabla_{i}\ln N$ is the acceleration. The coefficients $c_{m}^{\left(c_{0};d_{2},d_{3}\right)}$ and $\tilde{c}_{m}^{\left(c_{0};d_{2},d_{3}\right)}$ are generally functions of $t$ and $N$ without spatial derivatives.
Note the spatial derivatives of Ricci tensor, like $\nabla_{i}R_{jk}$, are not included in our model as they are higher order in $d$.

Without any specific fine tuning of the coefficients, in \cite{Gao:2014soa} it has been shown through a Hamiltonian analysis that the action (\ref{S_form}) propagates 3 DOF's in general, of which one is a scalar mode and two are tensor modes\footnote{In \cite{Gao:2014soa} only the parity-preserving Lagrangian is considered. While from the analysis it is clear that the presence of $\varepsilon_{ijk}$ would not change the constraint structure and thus the number of DOF's.}.
Therefore, in order to eliminate one of the three DOF'S, in particular, the scalar mode, additional conditions must be imposed.

For the action in (\ref{S_form}), the general conditions of propagating at most two tensorial degrees of freedom (TTDOF's) have been derived in \cite{Gao:2019twq}, which can be written as 
\begin{eqnarray}
	\mathcal{S}(\vec{x},\vec{y}) \approx 0 ,\qquad \mathcal{J}(\vec{x},\vec{y}) \approx 0,
\end{eqnarray}
where
\begin{eqnarray}
	\mathcal{S}\left(\vec{x},\vec{y}\right) & \coloneqq & \frac{\delta^{2}S}{\delta N\left(\vec{x}\right)\delta N\left(\vec{y}\right)}-\int\mathrm{d}^{3}x'\int\mathrm{d}^{3}y'\,N\left(\vec{x}'\right)\frac{\delta}{\delta N\left(\vec{x}\right)}\left(\frac{1}{N\left(\vec{x}'\right)}\frac{\delta S}{\delta K_{i'j'}\left(\vec{x}'\right)}\right)\\
	&  & \hspace{8em}\times\mathcal{G}_{i'j',k'l'}\left(\vec{x}',\vec{y}'\right)N\left(\vec{y}'\right)\frac{\delta}{\delta N\left(\vec{y}\right)}\left(\frac{1}{N\left(\vec{y}'\right)}\frac{\delta S}{\delta K_{i'j'}\left(\vec{y}'\right)}\right),
\end{eqnarray}
with $\mathcal{G}_{ij,kl}\left(\vec{x},\vec{y}\right)$ the inverse
of the Hessian with respect to $K_{ij}$ satisfying
\begin{equation}
	\int\mathrm{d}^{3}z\,\mathcal{G}_{ij,mn}\left(\vec{x},\vec{z}\right)\frac{\delta^{2}S}{\delta K_{mn}\left(\vec{z}\right)\delta K_{kl}\left(\vec{y}\right)}\equiv\frac{1}{2}\left(\delta_{i}^{k}\delta_{j}^{l}+\delta_{i}^{l}\delta_{j}^{k}\right)\delta^{3}\left(\vec{x}-\vec{y}\right),
\end{equation}
and
\begin{eqnarray}
	\mathcal{J}\left(\vec{x},\vec{y}\right) & \coloneqq & \int\mathrm{d}^{3}x'\int\mathrm{d}^{3}y'\int\mathrm{d}^{3}x''\int\mathrm{d}^{3}y''\frac{\delta C\left(\vec{x}\right)}{\delta K_{ij}\left(\vec{x}'\right)}\mathcal{G}_{i'j',k'l'}\left(\vec{x}',\vec{x}''\right)\nonumber \\
	&  & \qquad\times N\left(\vec{x}''\right)\frac{\delta^{2}S}{\delta h_{i'j'}\left(\vec{x}''\right)\delta K_{k'l'}\left(\vec{y}''\right)}\mathcal{G}_{k'l',kl}\left(\vec{y}'',\vec{y}'\right)\frac{\delta C\left(\vec{y}\right)}{\delta K_{ij}\left(\vec{y}'\right)}\nonumber \\
	&  & -\int\mathrm{d}^{3}x'\int\mathrm{d}^{3}y'\frac{\delta C'\left(\vec{x}\right)}{\delta K_{ij}\left(\vec{x}'\right)}\mathcal{G}_{ij,kl}\left(\vec{x}',\vec{y}'\right)N\left(\vec{y}'\right)\frac{\delta C\left(\vec{y}\right)}{\delta h_{kl}\left(\vec{y}'\right)}-\left(\vec{x}\leftrightarrow\vec{y}\right),
\end{eqnarray}
with
\begin{equation}
	C\left(\vec{x}\right)\coloneqq-\frac{\delta S}{\delta N\left(\vec{x}\right)}+\frac{1}{N\left(\vec{x}\right)}\frac{\delta S}{\delta K_{ij}\left(\vec{x}\right)}K_{ij}\left(\vec{x}\right).
\end{equation}
These two TTDOF conditions, which are dubbed the degenerate condition and the consistency condition, are the necessary and sufficient conditions for the action (\ref{S_form}) to propagate at most two DOF's.
When the lapse function becomes dynamical, the generalized conditions have also been derived in \cite{Lin:2020nro}.

Although these TTDOF conditions are general and conceptually simple, they are mathematically involved to be solved to yield concrete Lagrangians.
This is one of the motivations of this work to look for an alternative and more practical approach.

\section{The perturbative approach and degeneracy condition} \label{sec:pert_degen}

The unwanted degree of freedom, if not contained in the theory, will never show up at any order in perturbations around some background.
Thus one may tune the coefficients in the Lagrangian such that the unwanted scalar mode is eliminated order by order in perturbations.
Since there is a finite number of conditions in the non-perturbative sense, one will stop at some finite order and get the final Lagrangian in which the scalar mode is fully eliminated.
The perturbative approach can be a possible candidate method to bypass the mathematical difficulties in deal with the non-perturbative conditions got in a Hamiltonian analysis. 
For a class of SCG theories with the dynamical lapse function, this perturbative analysis has been used  to reduce the number of DOF's from four to three  \cite{Gao:2019lpz}. 
It was interesting that even at the cubic order in perturbations around a cosmological background, one could re-produce the fully non-perturbative conditions to eliminate the unwanted mode.

We consider perturbations around a Friedmann-Robertson-Walker (FRW) background. For our purpose, we focus on the scalar perturbations only.
After fixing the gauge freedom of the spatial diffeomorphism, the usual Arnowitt-Deser-Misner  (ADM) variables correspond to
\begin{eqnarray}
	N & = & \bar{N}e^{A},\label{lapse_xpn}\\
	N_{i} & = & \bar{N}a\,\partial_{i}B,\label{shift_xpn}\\
	h_{ij} & = & a^{2}e^{2\zeta}\delta_{ij},\label{sptmetric_xpn}
\end{eqnarray}
where $a=a(t)$ is the scale factor, $\bar{N} = \bar{N}(t)$ is background value of the lapse function.

Contrary to what one usually did in generally covariant theories, here we do not set $\bar{N} = 1$, since there is no time-reparametrization symmetry in our theory in general.
In particular, we assume that the Lagrangian depends on the lapse function $N$ explicitly while not on the time. 
As a result, generally the lapse function $N$ has a non-unity background value $\bar{N}(t)$. 
On the other hand, setting $\bar{N}=1$ implicitly redefines the time parameter $t$, which re-introduces the time-dependence of the Lagrangian.

The quadratic action for the scalar perturbations takes the form (we follow the notation in \cite{Gao:2019lpz})
	\begin{equation}	S_{2}\left[\zeta,A,B\right]=\int\!\mathrm{d}t\mathrm{d}^{3}x\,\bar{N}a^{3}\mathcal{L}_{2}.\label{S2_genstru}
	\end{equation}
The quadratic Lagrangian can be split into two parts
	\begin{equation}
		\mathcal{L}_{2}=\mathcal{L}_{2}^{(\mathrm{I})}+\mathcal{L}_{2}^{(\mathrm{II})},
	\end{equation}
in which $\mathcal{L}_{2}^{(\mathrm{I})}$ stands for terms relevant to counting the number of DOF's,
\begin{equation}
	\mathcal{L}_{2}^{(\mathrm{I})}=\dot{\zeta}\hat{\mathcal{C}}_{\dot{\zeta}\dot{\zeta}}\dot{\zeta}+\dot{\zeta}\hat{\mathcal{C}}_{\dot{\zeta}A}A+\dot{\zeta}\hat{\mathcal{C}}_{\dot{\zeta}B}B+A\hat{\mathcal{C}}_{AA}A+A\hat{\mathcal{C}}_{AB}B+B\hat{\mathcal{C}}_{BB}B,\label{Lag2_part1}
\end{equation}
and $\mathcal{L}_{2}^{(\mathrm{II})}$ stands for terms irrelevant to counting the number of DOF's.
\begin{equation}
	\mathcal{L}_{2}^{(\mathrm{II})}=\zeta\hat{\mathcal{C}}_{\zeta\zeta}\zeta+\zeta\hat{\mathcal{C}}_{\zeta A}A+\zeta\hat{\mathcal{C}}_{\zeta B}B.\label{Lag2_part2}
\end{equation}
In the above $\hat{\mathcal{C}}_{\dot{\zeta}\dot{\zeta}},\hat{\mathcal{C}}_{\dot{\zeta}A}$ etc. are time-dependent operators which may contain spatial derivatives.
Following \cite{Fujita:2015ymn}, throughout this paper we shall use the shorthand
\begin{equation}
	\dot{X}\equiv\frac{1}{\bar{N}}\frac{\partial X}{\partial t},\qquad f'\equiv\bar{N}\left.\frac{\partial f}{\partial N}\right|_{N=\bar{N}},\qquad f''\equiv\bar{N}^{2}\left.\frac{\partial^{2}f}{\partial N^{2}}\right|_{N=\bar{N}}. \label{notations}
\end{equation}

At this point, note the quadratic Lagrangian for the scalar modes contain no parity-violating term.
In other words, the parity-violating terms in (\ref{calL_3p}) and (\ref{calL_4p}) do not contribute to the quadratic order Lagrangian for the scalar modes and have nothing to do with eliminating the scalar modes at least at the linear order in perturbations.
Mathematically, this is simply because it is not possible to build a term quadratic in the scalar modes with $\varepsilon_{ijk} $.
If we go to higher order, the parity-violating terms do contribute to the scalar modes.

It is clear that in the quadratic action (\ref{S2_genstru}), $A$ and $B$ act as the auxiliary variables (i.e., without the  time derivatives). We may solve $A$ and $B$ formally from their equations of motion
	\begin{eqnarray}
		2\frac{1}{\bar{N}}\partial_{t}\left(a^{3}\hat{\mathcal{C}}_{\dot{\zeta}\dot{\zeta}}\dot{\zeta}\right)+\frac{1}{\bar{N}}\partial_{t}\left(a^{3}\hat{\mathcal{C}}_{\dot{\zeta}A}A\right)+\frac{1}{\bar{N}}\partial_{t}\left(a^{3}\hat{\mathcal{C}}_{\dot{\zeta}B}B\right)-2a^{3}\hat{\mathcal{C}}_{\zeta\zeta}\zeta-a^{3}\hat{\mathcal{C}}_{\zeta A}A-a^{3}\hat{\mathcal{C}}_{\zeta B}B & = & 0,\label{eom_zeta}\\
		2\hat{\mathcal{C}}_{AA}A+\hat{\mathcal{C}}_{AB}B+\hat{\mathcal{C}}_{\dot{\zeta}A}\dot{\zeta}+\hat{\mathcal{C}}_{\zeta A}\zeta & = & 0,\label{eom_A}\\
		\hat{\mathcal{C}}_{AB}A+2\hat{\mathcal{C}}_{BB}B+\hat{\mathcal{C}}_{\dot{\zeta}B}\dot{\zeta}+\hat{\mathcal{C}}_{\zeta B}\zeta & = & 0.\label{eom_B}
	\end{eqnarray}
The solutions for $A$ and $B$ can be formally written as
	\begin{equation} 
		A=\frac{\left(\frac{1}{2}\hat{\mathcal{C}}_{AB}\hat{\mathcal{C}}_{\dot{\zeta}B}-\hat{\mathcal{C}}_{BB}\hat{\mathcal{C}}_{\dot{\zeta}A}\right)\dot{\zeta}+\left(\frac{1}{2}\hat{\mathcal{C}}_{AB}\hat{\mathcal{C}}_{\zeta B}-\hat{\mathcal{C}}_{BB}\hat{\mathcal{C}}_{\zeta A}\right)\zeta}{2\hat{\mathcal{C}}_{AA}\hat{\mathcal{C}}_{BB}-\frac{1}{2}\hat{\mathcal{C}}_{AB}\hat{\mathcal{C}}_{AB}},\label{conssol_A}
	\end{equation}
	and
	\begin{equation}
		B=\frac{\left(\frac{1}{2}\hat{\mathcal{C}}_{AB}\hat{\mathcal{C}}_{\dot{\zeta}A}-\hat{\mathcal{C}}_{AA}\hat{\mathcal{C}}_{\dot{\zeta}B}\right)\dot{\zeta}+\left(\frac{1}{2}\hat{\mathcal{C}}_{AB}\hat{\mathcal{C}}_{\zeta A}-\hat{\mathcal{C}}_{AA}\hat{\mathcal{C}}_{\zeta B}\right)\zeta}{2\hat{\mathcal{C}}_{AA}\hat{\mathcal{C}}_{BB}-\frac{1}{2}\hat{\mathcal{C}}_{AB}\hat{\mathcal{C}}_{AB}}, \label{conssol_B}
	\end{equation}
with $2\hat{\mathcal{C}}_{AA}\hat{\mathcal{C}}_{BB}-\frac{1}{2}\hat{\mathcal{C}}_{AB}\hat{\mathcal{C}}_{AB} \neq 0$.
Since $\hat{\mathcal{C}}_{\dot{\zeta}\dot{\zeta}},\hat{\mathcal{C}}_{\dot{\zeta}A}$ etc. may contain spatial derivatives, the above solutions may be better understood in the Fourier space.
Plugging the above solutions into (\ref{eom_zeta}) yields the equation of motion for the single variable $\zeta$.
If in the equation of motion $\zeta$ acquires a second derivative term $\ddot{\zeta}$, $\zeta$ is dynamical.
Therefore in order to have no scalar mode propagating at the linear order, we have to ``kill'' the coefficient of $\ddot{\zeta}$ in its equation of motion.
After some manipulations, one find that this implies
	\begin{equation}
		\Delta \coloneqq 2\hat{\mathcal{C}}_{\dot{\zeta}\dot{\zeta}}\left(2\hat{\mathcal{C}}_{AA}\hat{\mathcal{C}}_{BB}-\frac{1}{2}\hat{\mathcal{C}}_{AB}^{2}\right)+\hat{\mathcal{C}}_{AB}\hat{\mathcal{C}}_{\dot{\zeta}A}\hat{\mathcal{C}}_{\dot{\zeta}B}-\hat{\mathcal{C}}_{BB}\hat{\mathcal{C}}_{\dot{\zeta}A}^{2}-\hat{\mathcal{C}}_{AA}\hat{\mathcal{C}}_{\dot{\zeta}B}^{2}=0.\label{degen_con}
	\end{equation}
We may refer to (\ref{degen_con}) the degeneracy condition.
The main task in this work is thus to use (\ref{degen_con}) as our starting point to find the conditions for the various coefficients $c_{m}^{\left(c_{0};d_{2},d_{3}\right)}$ and $\tilde{c}_{m}^{\left(c_{0};d_{2},d_{3}\right)}$ such that such that no scalar mode is propagating at the linear order in perturbations.

In the above, we derive the degeneracy condition (\ref{S2_genstru}) by solving the auxiliary variables and looking at the coefficient of the kinetic term in the effective Lagrangian for the single variable $\zeta$, which is the standard operation in calculating the cosmological perturbations.
We emphasize that it is not trivial to count the number of dynamical degree of freedom even for the quadratic Lagrangian in (\ref{S2_genstru}).
In the appendix \ref{app:clsmech}, we make a thorough analysis of a point particle model.


\section{Eliminate the scalar mode at the linear order} \label{sec:quad}

In this section, we shall find the conditions for the coefficients in the Lagrangian such that the degeneracy condition (\ref{S2_genstru}) is satisfied, and thus no scalar mode propagates at the linear order in perturbations.

\subsection{$d=2$}
We consider the model constructed by all the terms of $d=2$:
	\begin{equation} 
	S=\int\mathrm{d}t\mathrm{d}^{3}x\,N\sqrt{h}\left(\mathcal{L}^{\left(2\right)}-\Lambda\right), \label{S_d2}
	\end{equation}
where $\mathcal{L}^{(2)}$ is given in (\ref{calL_2}).
We have introduced a positive cosmological constant $\Lambda>0$ in order to have an expanding background solution. 
Equivalently, the above Lagrangian can be regarded as the linear combination of $\mathcal{L}^{(0)}$ and $\mathcal{L}^{(2)}$ with $c_{1}^{(0;0,0)}=\Lambda$.

Expanding the action to the first order in perturbations yields
	\begin{equation}
		S_{1}=\int\mathrm{d}t\mathrm{d}^{3}x\,\mathcal{L}_{1},
	\end{equation}
with
	\begin{eqnarray}
		\mathcal{L}_{1} & \simeq & \bar{N}a^{3}\left[-3H^{2}\left(b_{2}-b_{2}'\right)-\Lambda\right]A\nonumber \\
		&  & +3\bar{N}a^{3}\left[-\left(3H^{2}+2\dot{H}\right)b_{2}-\Lambda-2H\dot{b}_{2}\right]\zeta,\label{Lpert_1_d2}
	\end{eqnarray}
where we define
	\begin{equation}
		b_{2}\equiv c_{1}^{(0,2,0)}+3c_{3}^{(0,2,0)},\label{b2_def}
	\end{equation}
for short. The background equations of motion are determined by requiring $S_1 = 0$, which are
	\begin{eqnarray}
		-3H^{2}\left(b_{2}-b_{2}'\right)-\Lambda & = & 0,\label{bgeom_d2_A}\\
		-\left(3H^{2}+2\dot{H}\right)b_{2}-\Lambda-2H\dot{b}_{2} & = & 0,\label{bgeom_d2_zeta}
	\end{eqnarray}
for $A$ and $\zeta$, respectively.
At this point, keep in mind that $\dot{f}$ and $f'$ are defined
as in (\ref{notations}). The Hubble parameter is defined to be $H\coloneqq \frac{\dot{a}}{a}\equiv \frac{1}{\bar{N} a} \frac{\partial a}{\partial t}$.
From (\ref{bgeom_d2_A}) it is transparent that we get a generally expanding background only with a non-vanishing cosmological constant $\Lambda$.

Expanding the action to the second order in perturbations yields
	\begin{equation}
		S_{2}=\int\mathrm{d}t\mathrm{d}^{3}x\,\bar{N}a^{3}\mathcal{L}_{2}.
	\end{equation}
According to (\ref{S2_genstru}), the relevant coefficients are
	\begin{eqnarray}
		\hat{\mathcal{C}}_{\dot{\zeta}\dot{\zeta}}^{(2)} & = & 3b_{2},\nonumber \\
		\hat{\mathcal{C}}_{\dot{\zeta}A}^{(2)} & = & 6H\left(b_{2}'-b_{2}\right),\\
		\hat{\mathcal{C}}_{\dot{\zeta}B}^{(2)} & = & -2b_{2}\frac{\partial^{2}}{a},\\
		\hat{\mathcal{C}}_{AA}^{(2)} & = & \frac{3}{2}H^{2}\left(b_{2}''-2b_{2}'+2b_{2}\right)-\tilde{b}_{2}\frac{\partial^{2}}{a^{2}},\\
		\hat{\mathcal{C}}_{AB}^{(2)} & = & 2H\left(-b_{2}'+b_{2}\right)\frac{\partial^{2}}{a},\\
		\hat{\mathcal{C}}_{BB}^{(2)} & = & \frac{1}{3}\left(2w_{2}+b_{2}\right)\frac{\partial^{4}}{a^{2}},
	\end{eqnarray}
where we denote
	\begin{equation}
		\tilde{b}_{2}\coloneqq c_{2}^{(0,2,0)},\label{b2t_def}
	\end{equation}
and
	\begin{equation}
		w_{2}\coloneqq c_{1}^{(0,2,0)},
	\end{equation}
for short.
For later convenience, we also have
	\begin{eqnarray}
		\hat{\mathcal{C}}_{\zeta A}^{(2)} & = & -4\left(h_{2}'+h_{2}\right)\frac{\partial^{2}}{a^{2}},
	\end{eqnarray}
with
	\begin{equation}
		h_{2}\coloneqq c_{1}^{(1,0,0)},
	\end{equation}
for short, and
	\begin{equation}
		\hat{\mathcal{C}}_{\zeta B}^{(2)}=0.\label{calC_zB_2}
	\end{equation}
In the above we have made use of the background equations of motion (\ref{bgeom_d2_A}) and (\ref{bgeom_d2_zeta}) to eliminate $\Lambda$ and to simplify the expressions of the coefficients.
From (\ref{conssol_A}) and (\ref{conssol_B}), the solutions for
$A$ and $B$ are
\begin{equation}
	A=\frac{4\left[3a^{2}H(b_{2}-b_{2}')w_{2}\dot{\zeta}+(h_{2}+h_{2}')(b_{2}+2w_{2})\partial^{2}\zeta\right]}{3a^{2}H^{2}\left[-2(b_{2}')^{2}+b_{2}\left(2b_{2}'+b_{2}''\right)+2\left(2b_{2}-2b_{2}'+b_{2}''\right)w_{2}\right]-2\tilde{b}_{2}(b_{2}+2w_{2})\partial^{2}},\label{A_sol_d2}
\end{equation}
and
\begin{equation}
	B=\frac{3a\left\{ 4H(b_{2}-b_{2}')(h_{2}+h_{2}')\partial^{2}\zeta-3a^{2}H^{2}\left(-2(b_{2}')^{2}+b_{2}(2b_{2}'+b_{2}'')\right)\dot{\zeta}+2b_{2}\tilde{b}_{2}\partial^{2}\dot{\zeta}\right\} }{\left\{ -3a^{2}H^{2}\left[-2(b_{2}')^{2}+b_{2}\left(2b_{2}'+b_{2}''\right)+2\left(2b_{2}-2b_{2}'+b_{2}''\right)w_{2}\right]+2\tilde{b}_{2}(b_{2}+2w_{2})\partial^{2}\right\} \partial^{2}}.\label{B_sol_d2}
\end{equation}

By plugging the above results in (\ref{degen_con}), we get the degeneracy condition
	\begin{equation}
	\Delta^{(2)}=12H^{2}w_{2}\left[-2\left(b_{2}'\right)^{2}+b_{2}\left(2b_{2}'+b_{2}''\right)\right]\frac{\partial^{4}}{a^{2}}-8b_{2}\tilde{b}_{2}w_{2}\frac{\partial^{6}}{a^{4}}. \label{degen_d2}
	\end{equation}
At this point, note we need to require that $w_{2}\neq0$, otherwise there will be no kinetic term for the gravitational waves \cite{Gao:2019liu}. 
In order to have $\Delta^{(2)}=0$ so that there is no scalar mode propagating, one special solution is 
	\begin{equation}
		b_{2}\equiv c_{1}^{(0,2,0)}+3c_{3}^{(0,2,0)}=0.
	\end{equation}
However, this choice is conflict with the background equation of motion (\ref{bgeom_d2_A}).
Therefore we must have $b_{2} \neq 0$, which is also the case of GR.
Then first we have to require that
	\begin{equation}
		\tilde{b}_{2} \equiv c_{2}^{(0,2,0)} =0, \label{bt2_cond}
	\end{equation}
so that $\sim \partial^{6}$ term in (\ref{degen_d2}) vanishes, since we have assumed $w_{2} \neq 0$.
This indicates that the acceleration $a_{i}$ should not appear explicitly.
We also need to require
	\begin{equation}
		-2\left(b_{2}'\right)^{2}+b_{2}\left(2b_{2}'+b_{2}''\right)=0.\label{b2_diffeqn}
	\end{equation}
Note (\ref{b2_diffeqn}) must hold for any value of $\bar{N}$. Therefore (\ref{b2_diffeqn}) is regarded as a homogeneous differential equation for $b_{2}$ as a function of $N$, in which $b_{2}'$ and $b_{2}''$ are defined as in (\ref{notations}).
For later convenience, the solutions for $A$ and $B$ (\ref{A_sol_d2}) in (\ref{B_sol_d2})
under the conditions (\ref{bt2_cond}) and (\ref{b2_diffeqn}) get simplified to be
	\begin{equation}
		A=\frac{b_{2}}{H\left(b_{2}-b_{2}'\right)}\dot{\zeta}+\frac{b_{2}(h_{2}+h_{2}')(b_{2}+2w_{2})}{3a^{2}H^{2}\left(b_{2}-b_{2}'\right)^{2}w_{2}}\partial^{2}\zeta,\label{A_sol_d2_red}
	\end{equation}
and
	\begin{equation}
		B=-\frac{b_{2}\left(h_{2}+h_{2}'\right)}{aH\left(b_{2}-b_{2}'\right)w_{2}}\zeta,\label{B_sol_d2_red}
	\end{equation}
which involves no $\dot{\zeta}$. In (\ref{A_sol_d2_red}) and (\ref{B_sol_d2_red}) we have made use of (\ref{b2_diffeqn}) to replace $b_{2}''$ in terms of $b_2$ and $b_{2}'$.

The general solution for $b_{2}$ to (\ref{b2_diffeqn}) is
	\begin{equation}
		b_{2}=\frac{C_{1}N}{1+C_{2}N}, \label{b2_sol}
	\end{equation}
where $C_{1},C_{2}$ are two constants. This solution is also consistent with the analysis in \cite{Gao:2019twq} (see eqs. (110)-(111) therein). Obviously, $b_{2}=\mathrm{const.}$
is a trivial solution, which corresponds to the limit $C_{1},C_{2}\rightarrow\infty$
by keeping $\frac{C_{1}}{C_{2}}$ finite. 
To conclude, the Lagrangian 
	\begin{equation}
		\mathcal{L}^{\left(2\right)} = w_2 \hat{K}_{ij}\hat{K}^{ij}+\frac{1}{3}\frac{C_{1}N}{1+C_{2}N}K^{2} +h_2 R, \label{calL_2_fin}
	\end{equation}
with $w_2$ and $h_2$ being general functions of $N$, contains no dynamical scalar degree of freedom at linear order in a cosmological background. Here $\hat{K}_{ij}$ is the traceless part of $K_{ij}$ defined by
	\begin{equation}
		\hat{K}_{ij}\coloneqq K_{ij}-\frac{1}{3}Kh_{ij}.
	\end{equation}

With the form of the Lagrangian (\ref{calL_2_fin}), GR is a special case with the choice
	\begin{equation}
		c_{1}^{\left(0;2,0\right)}=c_{1}^{\left(1;0,0\right)}=1,
	\end{equation}
and thus corresponds to
	\begin{equation}
	\left|C_{1}\right|,\left|C_{2}\right|\rightarrow\infty,\qquad\text{keeping}\qquad\frac{C_{1}}{C_{2}}=-2.
	\end{equation}


\subsection{$d=3$}

Next we consider
	\begin{equation}
	S=\int\mathrm{d}t\mathrm{d}^{3}x\,N\sqrt{h}\left(\mathcal{L}^{\left(3\right)} - \Lambda\right),
	\end{equation}
where $\mathcal{L}^{(3)}$ is given in (\ref{calL_3}), and again we include a positive cosmological constant $\Lambda$ in order to have a non-vanishing $H$.
Expanding the action to the first order in perturbations yields
	\begin{equation}
		S_{1}=\int\mathrm{d}t\mathrm{d}^{3}x\,\mathcal{L}_{1},
	\end{equation}
with
	\begin{eqnarray}
		\mathcal{L}_{1} & = & \bar{N}a^{3}\left[3H^{3}\left(b_{3}'-2b_{3}\right)-\Lambda\right]A\nonumber \\
		&  & +3\bar{N}a^{3}\left[-6H\left(H^{2}+\dot{H}\right)b_{3}-\Lambda-3H^{2}\dot{b}_{3}\right]\zeta,\label{Lpert_1_d3}
	\end{eqnarray}
where we define
	\begin{equation}
		b_{3}\equiv c_{1}^{(0,3,0)}+3c_{3}^{(0,3,0)}+9c_{5}^{(0,3,0)}, \label{b3_def}
	\end{equation}
for short. Thus the background equations of motion are
	\begin{eqnarray}
		-3H^{3}\left(2b_{3}-b_{3}'\right)-\Lambda & = & 0,\label{bgeom_d3_A}\\
		-6H\left(H^{2}+\dot{H}\right)b_{3}-\Lambda-3H^{2}\dot{b}_{3} & = & 0,\label{bgeom_d3_zeta}
	\end{eqnarray}
for $A$ and $\zeta$, respectively. Note we must have $b_{3} \neq 0$ in order to make (\ref{bgeom_d3_A}) consistent with a non-vanishing $\Lambda$.

Expanding the action to the second order in perturbations yields
	\begin{equation}
		S_{2}=\int\mathrm{d}t\mathrm{d}^{3}x\,\bar{N}a^{3}\mathcal{L}_{2}.
	\end{equation}
According to (\ref{S2_genstru}), the relevant coefficients are
	\begin{eqnarray}
		\hat{\mathcal{C}}_{\dot{\zeta}\dot{\zeta}}^{(3)} & = & 9Hb_{3},\\
		\hat{\mathcal{C}}_{\dot{\zeta}A}^{(3)} & = & -9H^{2}\left(-b_{3}'+2b_{3}\right)+f_{3}\frac{\partial^{2}}{a^{2}},\\
		\hat{\mathcal{C}}_{\dot{\zeta}B}^{(3)} & = & -6Hb_{3}\frac{\partial^{2}}{a},\\
		\hat{\mathcal{C}}_{AA}^{(3)} & = & \frac{3}{2}H^{3}\left(b_{3}''-4b_{3}'+6b_{3}\right)+H\left(f_{3}'-\tilde{b}_{3}\right)\frac{\partial^{2}}{a^{2}},\\
		\hat{\mathcal{C}}_{AB}^{(3)} & = & 3H^{2}\left(-b_{3}'+2b_{3}\right)\frac{\partial^{2}}{a}-\tilde{f}_{3}\frac{\partial^{4}}{a^{3}},\\
		\hat{\mathcal{C}}_{BB}^{(3)} & = & H\left(2w_{3}+b_{3}\right)\frac{\partial^{4}}{a^{2}},
	\end{eqnarray}
where we denote
	\begin{eqnarray}
		f_{3} & \coloneqq & c_{1}^{(0,1,1)}+3c_{2}^{(0,1,1)},\\
		\tilde{b}_{3} & \coloneqq & c_{2}^{(0,3,0)}+3c_{4}^{(0,3,0)},\\
		\tilde{f}_{3} & \coloneqq & c_{1}^{(0,1,1)}+c_{2}^{(0,1,1)},\\
		w_{3} & \coloneqq & c_{1}^{(0,3,0)}+c_{3}^{(0,3,0)},
	\end{eqnarray}
for shorthand.
We have made use of the background equations of motion (\ref{bgeom_d3_A}) and (\ref{bgeom_d3_zeta}) to simply the coefficients.

After some manipulations, the degeneracy condition (\ref{degen_con}) now becomes
	\begin{eqnarray}
		\Delta^{(3)} & = & 54H^{5}w_{3}\left[-3\left(b_{3}'\right)^{2}+2b_{3}\left(2b_{3}'+b_{3}''\right)\right]\frac{\partial^{4}}{a^{2}}\nonumber \\
		&  & +36H^{3}w_{3}\left[-b_{3}'f_{3}+2b_{3}\left(f_{3}+f_{3}'-\tilde{b}_{3}\right)\right]\frac{\partial^{6}}{a^{4}}\nonumber \\
		&  & -H\left[b_{3}\left(f_{3}-3\tilde{f}_{3}\right)^{2}+2\left(f_{3}\right)^{2}w_{3}\right]\frac{\partial^{8}}{a^{6}}.
	\end{eqnarray}
Similar to the case of $d=2$, we require $w_{3}\neq0$, otherwise there will be no
gravitational waves \cite{Gao:2019liu}. Thus the degeneracy condition $\Delta^{(3)}=0$ yields a set of three equations
\begin{eqnarray}
	-3\left(b_{3}'\right)^{2}+2b_{3}\left(2b_{3}'+b_{3}''\right) & = & 0,\label{b3_diffeqn}\\
	-b_{3}'f_{3}+2b_{3}\left(f_{3}+f_{3}'-\tilde{b}_{3}\right) & = & 0, \label{bt3_eqn} \\
	b_{3}\left(f_{3}-3\tilde{f}_{3}\right)^{2}+2\left(f_{3}\right)^{2}w_{3} & = & 0. \label{f3ft3_eqn}
\end{eqnarray}
Recall that  there are nine free coefficients in the original Lagrangian $\mathcal{L}^{(3)}$ (\ref{calL_3}), which are subject to the above three equations in order to eliminate the scalar degree of freedom at the linear order.
We can solve $b_{3}$ from (\ref{b3_diffeqn}) to be
	\begin{equation}
		b_{3}=\frac{D_{1}N^{2}}{\left(1+D_{2}N\right)^{2}}, \label{b3_sol_d3}
	\end{equation}
with $D_{1},D_{2}$ being two constants.
At this point, note in order to make the background equation of motion (\ref{bgeom_d3_A}) consistent, which now reads
	\begin{equation}
		6H^{3}\frac{D_{1}D_{2}N^{3}}{\left(1+D_{2}N\right)^{3}}+\Lambda=0,
	\end{equation}
we have to require that (since $N>0$)
	\begin{equation}
		\frac{D_{1}D_{2}}{\left(1+D_{2}N\right)^{3}}<0.
	\end{equation}

By using the solution (\ref{b3_sol_d3}), we then solve $\tilde{b}_{3}$ from (\ref{bt3_eqn}) to be
\begin{equation}
	\tilde{b}_{3}=\frac{D_{2}N}{1+D_{2}N}f_{3}+f_{3}'.
\end{equation}

As for the last equation (\ref{f3ft3_eqn}), according to $f_{3}$ is vanishing or not, we discuss two cases.

\subsubsection{Case 1}

If $f_{3}=0$, since we assume $b_{3}\neq 0$, from (\ref{f3ft3_eqn})
we must also have $\tilde{f}_{3}=0$, which implies that
\begin{equation}
	c_{1}^{(0;1,1)}=0,\qquad c_{2}^{(0;1,1)}=0,
\end{equation}
and $w_{3}$ can be chosen freely (but non-vanishing) in general. 
In this case, there is no $\nabla a$ term in the Lagrangian.
As a result, $\tilde{b}_{3}=0$
and thus we may solve
\begin{equation}
	c_{4}^{(0;3,0)}=-\frac{1}{3}c_{2}^{(0;3,0)}.
\end{equation}

In this case, the Lagrangian is given by
\begin{eqnarray}
	\mathcal{L}^{\left(3\right),\mathrm{I}} & = & c_{1}^{\left(0;3,0\right)}\hat{K}_{ij}\hat{K}^{jk}\hat{K}_{k}^{i}+w_{3}\,\hat{K}_{ij}\hat{K}^{ij}K+\frac{1}{9}\frac{D_{1}N^{2}}{\left(1+D_{2}N\right)^{2}}K^{3}\nonumber \\
	&  & +c_{2}^{\left(0;3,0\right)}\hat{K}_{ij}a^{i}a^{j}+c_{1}^{\left(1;1,0\right)}R^{ij}K_{ij}+c_{2}^{\left(1;1,0\right)}RK, \label{calL_3_fin_1}
\end{eqnarray}
where the coefficients $c_{1}^{\left(0;3,0\right)}$, $w_{3}$ etc. are generally functions of $N$.

\subsubsection{Case 2}

If $f_{3}\neq0$, we can solve $w_{3}$ or more conveniently $c_{3}^{(0;3,0)}$
from (\ref{f3ft3_eqn}) to be
	\begin{equation}
	c_{3}^{(0;3,0)}=-c_{1}^{(0;3,0)}-2\frac{D_{1}N^{2}}{\left(1+D_{2}N\right)^{2}}\left(\frac{c_{1}^{(0;1,1)}}{f_{3}}\right)^{2},
	\end{equation}
As a result, by making use of (\ref{b3_def}) and (\ref{b3_sol_d3}), we may solve
	\begin{equation}
	c_{5}^{(0;3,0)}=\frac{2}{9}c_{1}^{(0;3,0)}+\frac{2}{3}\frac{D_{1}N^{2}}{\left(1+D_{2}N\right)^{2}}\left(\frac{c_{1}^{(0;1,1)}}{f_{3}}\right)^{2}+\frac{1}{9}\frac{D_{1}N^{2}}{\left(1+D_{2}N\right)^{2}}.
	\end{equation}

In this case the Lagrangian reduces to be
\begin{eqnarray}
	\mathcal{L}^{\left(3\right),\mathrm{II}} & = & c_{1}^{\left(0;3,0\right)}\hat{K}_{ij}\hat{K}^{jk}\hat{K}_{k}^{i}+\frac{1}{9}\frac{D_{1}N^{2}}{\left(1+D_{2}N\right)^{2}}K^{3}\nonumber \\
	&  & -2\frac{D_{1}N^{2}}{\left(1+D_{2}N\right)^{2}}\left(\frac{c_{1}^{(0;1,1)}}{f_{3}}\right)^{2}\hat{K}_{ij}\hat{K}^{ij}K\nonumber \\
	&  & +c_{2}^{\left(0;3,0\right)}\hat{K}_{ij}a^{i}a^{j}+\frac{1}{3}\left(\frac{D_{2}N}{1+D_{2}N}f_{3}+f_{3}'\right)Ka_{i}a^{i}\nonumber \\
	&  & +c_{1}^{\left(0;1,1\right)}\hat{K}_{ij}\nabla^{i}a^{j}+\frac{1}{3}f_{3}K\nabla_{i}a^{i}\nonumber \\
	&  & +c_{1}^{\left(1;1,0\right)}R^{ij}K_{ij}+c_{2}^{\left(1;1,0\right)}RK, \label{calL_3_fin_2}
\end{eqnarray}
which contains the spatial derivative terms of the acceleration $\nabla a$.

We thus conclude that (\ref{calL_3_fin_1}) and (\ref{calL_3_fin_2}) are two viable Lagrangians that do not propagate any scalar mode at the linear order in a cosmological background.

\subsection{$d=2$ with $d=3$}

In the above we have determined the viable Lagrangians when only $d=2$ or $d=3$ terms are present.
It should not be surprising that although the scalar mode is eliminated for $d=2$ and $d=3$ individually, the scalar mode will reappear if we naively combine them.
This also happens in the investigation of degenerate higher order scalar-tensor theories  \cite{Langlois:2018dxi,Langlois:2015cwa,BenAchour:2016fzp}.
Fortunately in our case, viable Lagrangians with the combination of $d=2$ and $d=3$ terms do exist, after imposing additional conditions on the coefficients.

We consider the combined Lagrangian 
\begin{equation}
	S=\int\mathrm{d}t\mathrm{d}^{3}x\,N\sqrt{h}\left(\mathcal{L}^{\left(2\right)}+\mathcal{L}^{\left(3\right)}-\Lambda\right),
\end{equation}
in which $\mathcal{L}^{\left(2\right)}$ and $\mathcal{L}^{\left(3\right)}$ are given in (\ref{calL_2}) and (\ref{calL_3}), respectively.

The analysis is completely parallel to the above.
Expanding the action to the first order in perturbations yields
\begin{equation}
	S_{1}=\int\mathrm{d}t\mathrm{d}^{3}x\,\mathcal{L}_{1},
\end{equation}
where it follows from (\ref{Lpert_1_d2}) and (\ref{Lpert_1_d3})
that
\begin{eqnarray}
	\mathcal{L}_{1} & = & \bar{N}a^{3}\left[-3H^{2}\left(b_{2}-b_{2}'\right)+3H^{3}\left(b_{3}'-2b_{3}\right)-\Lambda\right]A\nonumber \\
	&  & +3\bar{N}a^{3}\left[-\left(3H^{2}+2\dot{H}\right)b_{2}-2H\dot{b}_{2}-6H\left(H^{2}+\dot{H}\right)b_{3}-\Lambda-3H^{2}\dot{b}_{3}\right]\zeta.
\end{eqnarray}

The coefficients in the quadratic Lagrangian for the scalar modes read
\begin{eqnarray}
	\hat{\mathcal{C}}_{\dot{\zeta}\dot{\zeta}}^{(2)+(3)} & = & \hat{\mathcal{C}}_{\dot{\zeta}\dot{\zeta}}^{(2)}+\hat{\mathcal{C}}_{\dot{\zeta}\dot{\zeta}}^{(3)},\nonumber \\
	\hat{\mathcal{C}}_{\dot{\zeta}A}^{(2)+(3)} & = & \hat{\mathcal{C}}_{\dot{\zeta}A}^{(2)}+\hat{\mathcal{C}}_{\dot{\zeta}A}^{(3)},\\
	\hat{\mathcal{C}}_{\dot{\zeta}B}^{(2)+(3)} & = & \hat{\mathcal{C}}_{\dot{\zeta}B}^{(2)}+\hat{\mathcal{C}}_{\dot{\zeta}B}^{(3)},\\
	\hat{\mathcal{C}}_{AA}^{(2)+(3)} & = & \hat{\mathcal{C}}_{AA}^{(2)}+\hat{\mathcal{C}}_{AA}^{(3)},\\
	\hat{\mathcal{C}}_{AB}^{(2)+(3)} & = & \hat{\mathcal{C}}_{AB}^{(2)}+\hat{\mathcal{C}}_{AB}^{(3)},\\
	\hat{\mathcal{C}}_{BB}^{(2)+(3)} & = & \hat{\mathcal{C}}_{BB}^{(2)}+\hat{\mathcal{C}}_{BB}^{(3)}.
\end{eqnarray}

The degeneracy condition thus becomes
\begin{eqnarray}
	\Delta^{(2)+(3)} & = & \Big\{12H^{2}\left[-2(b_{2}')^{2}+b_{2}(2b_{2}'+b_{2}'')\right]w_{2}+54H^{5}\left[-3(b_{3}')^{2}+2b_{3}(2b_{3}'+b_{3}'')\right]w_{3}\nonumber \\
	&  & \quad+H^{3}\left[12\left(3b_{3}\left(b_{2}''+2b_{2}'\right)+b_{2}\left(2b_{3}'+b_{3}''\right)-6b_{2}'b_{3}'\right)w_{2}+36\left(-2(b_{2}')^{2}+b_{2}(2b_{2}'+b_{2}'')\right)w_{3}\right]\nonumber \\
	&  & +H^{4}\left[18\left(-3(b_{3}')^{2}+2b_{3}(2b_{3}'+b_{3}'')\right)w_{2}+36\left(3b_{3}\left(b_{2}''+2b_{2}'\right)+b_{2}\left(2b_{3}'+b_{3}''\right)-6b_{2}'b_{3}'\right)w_{3}\right]\Big\}\frac{\partial^{4}}{a^{2}}\nonumber \\
	&  & +\Big\{-8b_{2}\tilde{b}_{2}w_{2}+36H^{3}\left[-b_{3}'f_{3}+2b_{3}(f_{3}+f_{3}'-\tilde{b}_{3})\right]w_{3}\nonumber \\
	&  & \quad+H\left[8\left(-b_{2}'f_{3}-3b_{3}\tilde{b}_{2}+b_{2}(f_{3}+f_{3}'-\tilde{b}_{3})\right)w_{2}-24b_{2}\tilde{b}_{2}w_{3}\right]\nonumber \\
	&  & \quad+H^{2}\left[12\left(-b_{3}'f_{3}+2b_{3}(f_{3}+f_{3}'-\tilde{b}_{3})\right)w_{2}+24\left(-b_{2}'f_{3}-3b_{3}\tilde{b}_{2}+b_{2}(f_{3}+f_{3}'-\tilde{b}_{3})\right)w_{3}\right]\Big\}\frac{\partial^{6}}{a^{4}}\nonumber \\
	&  & +\left\{ -\frac{1}{3}\left(b_{2}(f_{3}-3\tilde{f}_{3})^{2}+2(f_{3})^{2}w_{2}\right)-H\left[b_{3}(f_{3}-3\tilde{f}_{3})^{2}+2(f_{3})^{2}w_{3}\right]\right\} \frac{\partial^{8}}{a^{6}}.
\end{eqnarray}
Note we have used the background equations of motion to simply the expressions for the coefficients.
We require that the degeneracy condition should be satisfied for any power of $\partial$ and $H$, thus we get the set of constraints
\begin{eqnarray}
	-2(b_{2}')^{2}+b_{2}\left(2b_{2}'+b_{2}''\right) & = & 0,\label{d2d3_eqn_1}\\
	-3(b_{3}')^{2}+2b_{3}\left(2b_{3}'+b_{3}''\right) & = & 0,\label{d2d3_eqn_2}\\
	3b_{3}\left(b_{2}''+2b_{2}'\right)+b_{2}\left(2b_{3}'+b_{3}''\right)-6b_{2}'b_{3}' & = & 0,\label{d2d3_eqn_3}\\
	\tilde{b}_{2} & = & 0,\label{d2d3_eqn_4}\\
	-b_{3}'f_{3}+2b_{3}(f_{3}+f_{3}'-\tilde{b}_{3}) & = & 0,\label{d2d3_eqn_5}\\
	-b_{2}'f_{3}-3b_{3}\tilde{b}_{2}+b_{2}(f_{3}+f_{3}'-\tilde{b}_{3}) & = & 0,\label{d2d3_eqn_6}\\
	b_{2}(f_{3}-3\tilde{f}_{3})^{2}+2(f_{3})^{2}w_{2} & = & 0,\label{d2d3_eqn_7}\\
	b_{3}(f_{3}-3\tilde{f}_{3})^{2}+2(f_{3})^{2}w_{3} & = & 0.\label{d2d3_eqn_8}
\end{eqnarray}
Compared with the constraints in $d=2$ and $d=3$ respectively, we find that additional constraints should be imposed in the combined case to ensure the degeneracy. 

First of all, (\ref{d2d3_eqn_1}) and (\ref{d2d3_eqn_2}) are exactly
(\ref{b2_diffeqn}) and (\ref{b3_diffeqn}) in the case of $d=2$
and $d=3$ individually. 
The general solutions for $b_{2}$ and $b_{3}$ are given in (\ref{b2_sol}) and (\ref{b3_sol_d3}), respectively, from which we may solve 
	\begin{equation}
		c_{3}^{(0,2,0)}=-\frac{1}{3}c_{1}^{(0,2,0)}+\frac{1}{3}\frac{C_{1}N}{1+C_{2}N},
	\end{equation}
and
	\begin{equation}
		c_{5}^{(0;3,0)}=-\frac{1}{9}c_{1}^{(0;3,0)}-\frac{1}{3}c_{3}^{(0;3,0)}+\frac{1}{9}\frac{D_{1}N^{2}}{\left(1+D_{2}N\right)^{2}}.\label{c0305_sol_d2d3_ori}
	\end{equation}
If both (\ref{d2d3_eqn_1}) and (\ref{d2d3_eqn_2})
are satisfied, (\ref{d2d3_eqn_3}) reduces to be
	\begin{equation}
		\frac{3}{2b_{2}b_{3}}\left(2b_{3}b_{2}'-b_{2}b_{3}'\right)^{2}=0,
	\end{equation}
which yields a constraint between $b_{2}$ and $b_{3}$. By plugging the solutions (\ref{b2_sol}) and (\ref{b3_sol_d3}), we may solve 
	\begin{equation}
		C_{2}=D_{2},
	\end{equation}	
which implies that $b_{3}$ is determined by $b_{2}$ through
	\begin{equation}
		b_{3}=\frac{D_{1}}{C_{1}^{2}}b_{2}^{2}.\label{b3_sol_d2d3}
	\end{equation}
As a result, (\ref{c0305_sol_d2d3_ori}) reduces to be
	\begin{equation}		c_{5}^{(0;3,0)}=-\frac{1}{9}c_{1}^{(0;3,0)}-\frac{1}{3}c_{3}^{(0;3,0)}+\frac{1}{9}\frac{D_{1}N^{2}}{\left(1+C_{2}N\right)^{2}}.\label{c0305_sol_d2d3}
	\end{equation}

(\ref{d2d3_eqn_4}) implies
\begin{equation}
	c_{2}^{(0,2,0)}=0.
\end{equation}
With (\ref{b3_sol_d2d3}), the left-hand-side of (\ref{d2d3_eqn_5})
becomes
\begin{equation}
	\mathrm{L.H.S.}=2\frac{D_{1}}{C_{1}^{2}}b_{2}\left[-b_{2}'f_{3}+b_{2}\left(f_{3}+f_{3}'-\tilde{b}_{3}\right)\right],
\end{equation}
while due to (\ref{d2d3_eqn_4}), the left-hand-side of (\ref{d2d3_eqn_6})
becomes
\begin{equation}
	\mathrm{L.H.S.}\equiv-b_{2}'f_{3}+b_{2}\left(f_{3}+f_{3}'-\tilde{b}_{3}\right),
\end{equation}
which is proportional to (\ref{d2d3_eqn_5}). Generally, we look for
solutions with $b_{2}\neq0$ (since GR belongs to the case), thus (\ref{d2d3_eqn_5})
and (\ref{d2d3_eqn_6}) are satisfied only if
\begin{equation}
	-b_{2}'f_{3}+b_{2}\left(f_{3}+f_{3}'-\tilde{b}_{3}\right)=0.
\end{equation}
We thus solve $\tilde{b}_{3}$ to be
\[
\tilde{b}_{3}=\frac{C_{2}N}{1+C_{2}N}f_{3}+f_{3}',
\]
which also implies
\begin{equation}
	c_{4}^{(0;3,0)}=-\frac{1}{3}c_{2}^{(0;3,0)}+\frac{1}{3}\left(\frac{C_{2}N}{1+C_{2}N}f_{3}+f_{3}'\right).\label{c0304_sol_d2d3_ori}
\end{equation}

In order to make (\ref{d2d3_eqn_7}) and (\ref{d2d3_eqn_8}) have
solutions, $f_{3}$ and $f_{3}-3\tilde{f}_{3}$ must be either non-vanishing
or vanishing simultaneously. Therefore we have two cases.

\subsubsection{Case 1}

If $f_{3}=0$ and $f_{3}-3\tilde{f}_{3}=0$ (and thus $\tilde{f}_{3}=0$),
both (\ref{d2d3_eqn_7}) and (\ref{d2d3_eqn_8}) are automatically satisfied.
There is no restriction on $b_{2},b_{3},w_{2},w_{3}$. In this case,
since $f_{3}=\tilde{f}_{3}=0$ we have
\begin{equation}
	c_{1}^{(0;1,1)}=0,\qquad c_{2}^{(0;1,1)}=0.
\end{equation}
As a result, (\ref{c0304_sol_d2d3_ori}) reduces to be
\begin{equation}
	c_{4}^{(0;3,0)}=-\frac{1}{3}c_{2}^{(0;3,0)}.
\end{equation}

In this case, the Lagrangian is given by
\begin{eqnarray}
	\mathcal{L}^{(2)+(3),\mathrm{I}} & = & c_{1}^{\left(0;2,0\right)}\hat{K}_{ij}\hat{K}^{ij}+\frac{1}{3}\frac{C_{1}N}{1+C_{2}N}K^{2}+c_{1}^{\left(1;0,0\right)}R\nonumber \\
	&  & +c_{1}^{\left(0;3,0\right)}\hat{K}_{ij}\hat{K}^{jk}\hat{K}_{k}^{i}+w_{3}\,\hat{K}_{ij}\hat{K}^{ij}K+\frac{1}{9}\frac{D_{1}N^{2}}{\left(1+C_{2}N\right)^{2}}K^{3}\nonumber \\
	&  & +c_{2}^{\left(0;3,0\right)}\hat{K}_{ij}a^{i}a^{j}+c_{1}^{\left(1;1,0\right)}R^{ij}K_{ij}+c_{2}^{\left(1;1,0\right)}RK, \label{calL_23_fin_1}
\end{eqnarray}
which contains no spatial derivative of the acceleration $\nabla a$.

\subsubsection{Case 2}

If $f_{3}\neq0$ and $f_{3}-3\tilde{f}_{3}\neq0$, from (\ref{d2d3_eqn_7})
we solve
\begin{equation}
	w_{2}=-\frac{1}{2}b_{2}\left(1-3\frac{\tilde{f}_{3}}{f_{3}}\right)^{2},
\end{equation}
i.e.,
\begin{equation}
	c_{1}^{(0,2,0)}=-2\frac{C_{1}N}{1+C_{2}N}\left(\frac{c_{1}^{(0;1,1)}}{f_{3}}\right)^{2},
\end{equation}
and from (\ref{d2d3_eqn_8}) we solve
\begin{equation}
	w_{3}=-\frac{1}{2}b_{3}\left(1-3\frac{\tilde{f}_{3}}{f_{3}}\right)^{2},
\end{equation}
i.e.,
\begin{equation}
	c_{3}^{(0;3,0)}=-c_{1}^{(0;3,0)}-2\frac{D_{1}N^{2}}{\left(1+C_{2}N\right)^{2}}\left(\frac{c_{1}^{(0;1,1)}}{f_{3}}\right)^{2}.
\end{equation}
As a result, (\ref{c0305_sol_d2d3}) reduces to be
\begin{equation}
	c_{5}^{(0;3,0)}=\frac{2}{9}c_{1}^{(0;3,0)}+\frac{1}{9}\frac{D_{1}N^{2}}{\left(1+C_{2}N\right)^{2}}\left[1+6\left(\frac{c_{1}^{(0;1,1)}}{f_{3}}\right)^{2}\right].
\end{equation}

In this case, the Lagrangian is given by
\begin{eqnarray}
	\mathcal{L}^{(2)+(3),\mathrm{II}} & = & -2\frac{C_{1}N}{1+C_{2}N}\left(\frac{c_{1}^{(0;1,1)}}{f_{3}}\right)^{2}\hat{K}_{ij}\hat{K}^{ij}+\frac{1}{3}\frac{C_{1}N}{1+C_{2}N}K^{2}+c_{1}^{\left(1;0,0\right)}R\nonumber \\
	&  & +c_{1}^{\left(0;3,0\right)}\hat{K}_{ij}\hat{K}^{jk}\hat{K}_{k}^{i}-2\frac{D_{1}N^{2}}{\left(1+C_{2}N\right)^{2}}\left(\frac{c_{1}^{(0;1,1)}}{f_{3}}\right)^{2}\hat{K}_{ij}\hat{K}^{ij}K+\frac{1}{9}\frac{D_{1}N^{2}}{\left(1+C_{2}N\right)^{2}}K^{3}\nonumber \\
	&  & +c_{2}^{\left(0;3,0\right)}\hat{K}_{ij}a^{i}a^{j}+\frac{1}{3}\left(\frac{C_{2}N}{1+C_{2}N}f_{3}+f_{3}'\right)Ka_{i}a^{i}\nonumber \\
	&  & +c_{1}^{\left(0;1,1\right)}\hat{K}_{ij}\nabla^{i}a^{j}+\frac{1}{3}f_{3}K\nabla_{i}a^{i}\nonumber \\
	&  & +c_{1}^{\left(1;1,0\right)}R^{ij}K_{ij}+c_{2}^{\left(1;1,0\right)}RK, \label{calL_23_fin_2}
\end{eqnarray}
which contains spatial derivatives of the acceleration $\nabla a$.

We conclude that (\ref{calL_23_fin_1}) and (\ref{calL_23_fin_2}) are two viable Lagrangians in which both $d=2$ and $d=3$ terms are present, and propagate no scalar mode at the linear order in perturbations around a cosmological background.

\subsection{$d=4$}

Now we consider the most involved case of $d=4$.
The action is 
\begin{equation}
	S=\int\mathrm{d}t\mathrm{d}^{3}x\,N\sqrt{h}\left(\mathcal{L}^{\left(4\right)}+\Lambda\right),
\end{equation}
with $\mathcal{L}^{\left(4\right)}$ given in (\ref{calL_4}).
Expanding the action to the first order in perturbations yields
\begin{equation}
	S_{1}=\int\mathrm{d}t\mathrm{d}^{3}x\,\mathcal{L}_{1},
\end{equation}
with
\begin{eqnarray}
	\mathcal{L}_{1} & = & \bar{N}a^{3}\left[-9H^{4}\left(-b_{4}'+3b_{4}\right)+\Lambda\right]A\nonumber \\
	&  & +3\bar{N}a^{3}\left[-9H^{2}\left(3H^{2}+4\dot{H}\right)b_{4}+\Lambda-12H^{3}\dot{b}_{4}\right]\zeta,
\end{eqnarray}
where we denote
\begin{equation}
	b_{4}\coloneqq c_{2}^{(0,4,0)}+c_{4}^{(0,4,0)}+3c_{7}^{(0,4,0)}+9c_{9}^{(0,4,0)},
\end{equation}
for short. The background equations of motion are
\begin{eqnarray}
	-9H^{4}\left(-b_{4}'+3b_{4}\right)+\Lambda & = & 0,\\
	-9H^{2}\left(3H^{2}+4\dot{H}\right)b_{4}+\Lambda-12H^{3}\dot{b}_{4} & = & 0,
\end{eqnarray}
for $A$ and $\zeta$, respectively.

The relevant coefficients in the quadratic Lagrangian for the scalar modes are
\begin{eqnarray}
	\hat{\mathcal{C}}_{\dot{\zeta}\dot{\zeta}}^{(4)} & = & 54H^{2}b_{4}-d_{4}\frac{\partial^{2}}{a^{2}},\\
	\hat{\mathcal{C}}_{\dot{\zeta}A}^{(4)} & = & -36H^{3}\left(-b_{4}'+3b_{4}\right)+H\left(2d_{4}+2f_{4}-\tilde{f}_{4}\right)\frac{\partial^{2}}{a^{2}},\\
	\hat{\mathcal{C}}_{\dot{\zeta}B}^{(4)} & = & -36H^{2}b_{4}\frac{\partial^{2}}{a}+2\tilde{d}_{4}\frac{\partial^{4}}{a^{3}},\\
	\hat{\mathcal{C}}_{AA}^{(4)} & = & \frac{9}{2}H^{4}\left(b_{4}''-6b_{4}'+12b_{4}\right)\nonumber \\
	&  & +H^{2}\left(f_{4}'-d_{4}-\tilde{b}_{4}-f_{4}+\tilde{f}_{4}\right)\frac{\partial^{2}}{a^{2}}+\bar{d}_{4}\frac{\partial^{4}}{a^{4}},\\
	\hat{\mathcal{C}}_{AB}^{(4)} & = & 12H^{3}\left(-b_{4}'+3b_{4}\right)\frac{\partial^{2}}{a}+H\left(-2\tilde{d}_{4}-\hat{f}_{4}\right)\frac{\partial^{4}}{a^{3}},\\
	\hat{\mathcal{C}}_{BB}^{(4)} & = & 2H^{2}\left(w_{4}+3b_{4}\right)\frac{\partial^{4}}{a^{2}}-\hat{d}_{4}\frac{\partial^{6}}{a^{4}},
\end{eqnarray}
where we denote
\begin{eqnarray}
	d_{4} & \coloneqq & 3c_{1}^{(0,0,2)}+c_{2}^{(0,0,2)}+3c_{3}^{(0,0,2)}+9c_{4}^{(0,0,2)},\\
	f_{4} & \coloneqq & c_{1}^{(0,2,1)}+3c_{4}^{(0,2,1)}+3c_{5}^{(0,2,1)}+9c_{7}^{(0,2,1)},\\
	\tilde{f}_{4} & \coloneqq & c_{2}^{(0,2,1)}+3c_{3}^{(0,2,1)}.\\
	\tilde{d}_{4} & \coloneqq & c_{1}^{(0,0,2)}+c_{2}^{(0,0,2)}+2c_{3}^{(0,0,2)}+3c_{4}^{(0,0,2)}.\\
	\tilde{b}_{4} & \coloneqq & c_{1}^{(0,4,0)}+3c_{3}^{(0,4,0)}+3c_{5}^{(0,4,0)}+9c_{8}^{(0,4,0)},\\
	\bar{d}_{4} & \coloneqq & c_{5}^{(0,0,2)}+c_{6}^{(0,0,2)}.\\
	\hat{f}_{4} & \coloneqq & 2c_{1}^{(0,2,1)}-c_{2}^{(0,2,1)}-c_{3}^{(0,2,1)}+4c_{4}^{(0,2,1)}+2c_{5}^{(0,2,1)}+6c_{7}^{(0,2,1)}.\\
	w_{4} & \coloneqq & 3c_{2}^{(0,4,0)}+2c_{4}^{(0,4,0)}+3c_{7}^{(0,4,0)},\\
	\hat{d}_{4} & \coloneqq & c_{1}^{(0,0,2)}+c_{2}^{(0,0,2)}+c_{3}^{(0,0,2)}+c_{4}^{(0,0,2)},
\end{eqnarray} for shorthand.

After some manipulation, the degeneracy condition (\ref{degen_con}) is found to be
\begin{equation}
	\Delta^{(4)}=H^{8}\Delta_{4}^{(4)}\frac{\partial^{4}}{a^{2}}+H^{6}\Delta_{6}^{(4)}\frac{\partial^{6}}{a^{4}}+H^{4}\Delta_{8}^{(4)}\frac{\partial^{8}}{a^{6}}+H^{2}\Delta_{10}^{(4)}\frac{\partial^{10}}{a^{8}}+\Delta_{12}^{(4)}\frac{\partial^{12}}{a^{10}},
\end{equation}
where
\begin{equation}
	\Delta_{4}^{(4)}=648\left[-4\left(b_{4}'\right)^{2}+3b_{4}\left(2b_{4}'+b_{4}''\right)\right]w_{4},\label{Delta_4_4}
\end{equation}
\begin{eqnarray}
	\Delta_{6}^{(4)} & = & +36\left[-\left(2b_{4}'+b_{4}''\right)d_{4}+12b_{4}\left(f_{4}+f_{4}'-\tilde{b}_{4}\right)+4b_{4}'\left(-2f_{4}+\tilde{f}_{4}\right)\right]w_{4}\nonumber \\
	&  & +36\left[4(b_{4}')^{2}-3b_{4}(2b_{4}'+b_{4}'')\right]\left(d_{4}+9\hat{d}_{4}-6\tilde{d}_{4}\right),\label{Delta_4_6}
\end{eqnarray}
\begin{eqnarray}
	\Delta_{8}^{(4)} & = & -18\left(2b_{4}'+b_{4}''\right)\left(\tilde{d}_{4}\right)^{2}+24d_{4}\left[-b_{4}\left(f_{4}+f_{4}'-\tilde{b}_{4}\right)+b_{4}'\hat{f}_{4}\right]\nonumber \\
	&  & -6b_{4}\left(-2f_{4}+3\hat{f}_{4}+\tilde{f}_{4}\right)^{2}+24\tilde{d}_{4}\left[6b_{4}\left(f_{4}+f_{4}'-\tilde{b}_{4}\right)+b_{4}'\left(-2f_{4}-3\hat{f}_{4}+\tilde{f}_{4}\right)\right]\nonumber \\
	&  & +18\hat{d}_{4}\left[\left(2b_{4}'+b_{4}''\right)d_{4}-12b_{4}\left(f_{4}+f_{4}'-\tilde{b}_{4}\right)-4b_{4}'\left(-2f_{4}+\tilde{f}_{4}\right)\right]\nonumber \\
	&  & +2\left[-4d_{4}\left(f_{4}+f_{4}'-\tilde{b}_{4}\right)+216b_{4}\bar{d}_{4}-\left(-2f_{4}+\tilde{f}_{4}\right)^{2}\right]w_{4},\label{Delta_4_8}
\end{eqnarray}
\begin{eqnarray}
	\Delta_{10}^{(4)} & = & -4\left(f_{4}+f_{4}'-\tilde{b}_{4}\right)\left(\tilde{d}_{4}\right)^{2}+d_{4}\left[-24b_{4}\bar{d}_{4}+\left(\hat{f}_{4}\right)^{2}\right]+2\tilde{d}_{4}\left[72b_{4}\bar{d}_{4}+\hat{f}_{4}\left(-2f_{4}+\tilde{f}_{4}\right)\right]\nonumber \\
	&  & +\hat{d}_{4}\left[4d_{4}\left(f_{4}+f_{4}'-\tilde{b}_{4}\right)-216b_{4}\bar{d}_{4}+\left(-2f_{4}+\tilde{f}_{4}\right)^{2}\right]-8d_{4}\bar{d}_{4}w_{4},\label{Delta_4_10}
\end{eqnarray}
and
\begin{equation}
	\Delta_{12}^{(4)}=4\bar{d}_{4}\left[d_{4}\hat{d}_{4}-(\tilde{d}_{4})^{2}\right].\label{Delta_4_12}
\end{equation}
It is interesting that $\tilde{b}_{4}$ always arises in terms of
$\left(f_{4}+f_{4}'-\tilde{b}_{4}\right)$, and $\tilde{f}_{4}$ always
arises in terms of $\left(-2f_{4}+\tilde{f}_{4}\right)$.

Following the analysis in the above, we may solve the coefficients such that $\Delta^{(4)}=0$.
For example, from (\ref{Delta_4_4}) and (\ref{Delta_4_6}) we have two constraints
\begin{equation}
	-4\left(b_{4}'\right)^{2}+3b_{4}\left(2b_{4}'+b_{4}''\right)=0,
\end{equation}
and
\begin{equation}
	-\left(2b_{4}'+b_{4}''\right)d_{4}+12b_{4}\left(f_{4}+f_{4}'-\tilde{b}_{4}\right)+4b_{4}'\left(-2f_{4}+\tilde{f}_{4}\right)=0,
\end{equation}
combining which yields
\begin{equation}
	-\left(b_{4}'\right)^{2}d_{4}+9\left(b_{4}\right)^{2}\left(f_{4}+f_{4}'-\tilde{b}_{4}\right)+3b_{4}b_{4}'\left(-2f_{4}+\tilde{f}_{4}\right)=0.
\end{equation}
The full treatment of the case of $d=4$, however, is involved and out of the scope of the present work.

\section{Eliminate the scalar mode at the second order: $d=2$} \label{sec:cub}

In the previous section, we have eliminate the scalar mode at linear order in perturbations by making use of the degeneracy condition (\ref{degen_con}).
Clearly the conditions for the coefficients derived in the above are merely necessary conditions, which means that the scalar mode will reappear if we go to higher orders.
Thus one need to find the conditions for the coefficients such that the scalar mode is eliminated order by order.
In this section we take the case of $d=2$ as an illustrative example.

Expanding the action (\ref{S_d2}) to the cubic order in perturbations yields
	\begin{eqnarray}
		\mathcal{L}_{3} & = & 9\bar{N}a^{3}b_{2}\zeta\dot{\zeta}^{2}-\bar{N}a^{2}Hb_{2}\zeta^{2}\partial^{2}B-2\bar{N}ah_{2}\zeta^{2}\partial^{2}\zeta+a\bar{N}\tilde{b}_{2}\zeta\partial_{i}A\partial^{i}A-2a^{2}H\bar{N}b_{2}\zeta\partial_{i}\zeta\partial^{i}B\nonumber \\
		&  & -2a\bar{N}h_{2}\zeta\partial_{i}\zeta\partial^{i}\zeta+\frac{1}{3}a\bar{N}\left(-b_{2}+w_{2}\right)\zeta\left(\partial^{2}B\right)^{2}+\frac{2}{3}a\bar{N}\left(b_{2}+2w_{2}\right)\partial_{i}\zeta\partial^{i}B\partial^{2}B\nonumber \\
		&  & -2a\bar{N}w_{2}\partial_{i}\partial_{j}B\partial^{i}B\partial^{j}\zeta-2a\bar{N}w_{2}\partial^{i}B\partial_{j}\partial_{i}B\partial^{j}\zeta-a\bar{N}w_{2}\zeta\partial_{j}\partial_{i}B\partial^{j}\partial^{i}B\nonumber \\
		&  & +27\bar{N}a^{3}Hb_{2}\zeta^{2}\dot{\zeta}-2a^{2}b_{2}\zeta\partial^{2}B\partial_{t}\zeta-2\bar{N}a^{2}b_{2}\partial_{i}\zeta\partial^{i}B\dot{\zeta}+2a^{2}H\bar{N}\left(b_{2}-b_{2}'\right)A\zeta\partial^{2}B\nonumber \\
		&  & +2a^{2}H\bar{N}\left(b_{2}-b_{2}'\right)A\partial_{i}\zeta\partial^{i}B-18a^{3}HA\zeta\partial_{t}\zeta\left(b_{2}-b_{2}'\right)+2\bar{N}a^{2}\left(b_{2}-b_{2}'\right)A\partial^{2}B\dot{\zeta}\nonumber \\
		&  & -\frac{27}{2}\bar{N}a^{3}H^{2}\left(-2b_{2}+b_{2}'\right)\zeta^{3}+3\bar{N}a^{3}\left(-b_{2}+b_{2}'\right)A\dot{\zeta}^{2}+a\bar{N}\left(\tilde{b}_{2}+\tilde{b}_{2}'\right)A\partial_{i}A\partial^{i}A\nonumber \\
		&  & -4a\bar{N}\left(h_{2}+h_{2}'\right)A\zeta\partial^{2}\zeta-2a\bar{N}\left(h_{2}+h_{2}'\right)A\partial_{i}\zeta\partial^{i}\zeta+\frac{1}{3}a\bar{N}\left(-b_{2}+w_{2}+b_{2}'-w_{2}'\right)A\left(\partial^{2}B\right)^{2}\nonumber \\
		&  & +a\bar{N}\left(-w_{2}+w_{2}'\right)A\partial_{j}\partial_{i}B\partial^{j}\partial^{i}B+\frac{9}{2}\bar{N}a^{3}H^{2}\left(2b_{2}-2b_{2}'+b_{2}''\right)A^{2}\zeta\nonumber \\
		&  & -\bar{N}a^{2}H\left(b_{2}-b_{2}'+b_{2}''\right)A^{2}\partial^{2}B+3\bar{N}a^{3}H\left(b_{2}-b_{2}'+b_{2}''\right)A^{2}\dot{\zeta}\nonumber \\
		&  & -2\bar{N}a\left(h_{2}+3h_{2}'+h_{2}''\right)A^{2}\partial^{2}\zeta+\frac{1}{2}\bar{N}a^{3}H^{2}b_{2}^{(3)}A^{3},
	\end{eqnarray}
where we have used the background equations of motion to simply the coefficients.
No integration by parts has been performed at this point.

By making use of the degeneracy conditions (\ref{bt2_cond}) and (\ref{b2_diffeqn}),
and plugging the solutions for $A$ and $B$ in (\ref{A_sol_d2_red}) and (\ref{B_sol_d2_red}), after some manipulations, we get the induced cubic action $S_{3}[\zeta]$ for the single variable $\zeta$.
We tend not to present the full and explicit expression of $S_{3}[\zeta]$ due to its length. 
We pay special attention to the terms which are relevant to eliminating the scalar mode (i.e., $\zeta$), and have the following observations.
\begin{itemize}
	\item First, we found that there is no $\dot{\zeta}^{3}$, i.e., there is no terms with 3 time derivatives.
	\item Second, there is one ``dangerous'' term with two time derivatives:
	\begin{equation}
		-2\bar{N}a\frac{b_{2}^{2}\left(2h_{2}'+h_{2}''\right)}{H^{2}\left(b_{2}-b_{2}'\right)^{2}}\dot{\zeta}^{2}\partial^{2}\zeta,
	\end{equation}
	thus we need to require (since $b_{2}\neq0$)
	\begin{equation}
		2h_{2}'+h_{2}''=0,
	\end{equation}
	with $b_{2}-b_{2}'\neq0$.
	The general solution for $h_2$ is
	\begin{equation}
		h_{2}\left(N\right)=C_{3}+\frac{C_{4}}{N} , \label{h2_sol}
	\end{equation}
	where $C_3,C_4$ are constants.
	\item Third, there is one ``dangerous'' term with one time derivative:
	\begin{equation}
		\frac{2b_{2}\left(h_{2}+h_{2}'\right)^{2}\left[-2b_{2}w_{2}^{2}+2w_{2}^{2}b_{2}'+b_{2}^{2}\left(-w_{2}+w_{2}'\right)\right]}{3aH^{3}w_{2}^{2}\bigl(b_{2}-b_{2}'\bigr)^{3}}\bar{N}\dot{\zeta}\left(\partial^{2}\zeta\right)^{2},
	\end{equation}
	which implies that $w_2$ must be related to $b_2$ by
	\begin{equation}
		\frac{1}{w_{2}^{2}}\left(w_{2}-w_{2}'\right)=-2\frac{1}{b_{2}^{2}}\left(b_{2}-b_{2}'\right).
	\end{equation}
	With the solution for $b_2$ in (\ref{b2_sol}), the general solution for $w_{2}$ is
	\begin{equation}
		w_{2}=\frac{C_{5} C_{1}N}{1-2C_{5} C_{2}N}, \label{w2_sol}
	\end{equation}
	where $C_5$ is another constant.
\end{itemize}

By combining (\ref{bt2_cond}), (\ref{b2_sol}), (\ref{h2_sol})
and (\ref{w2_sol}), the Lagrangian (\ref{calL_2_fin}) is further reduced to be
	\begin{eqnarray}
		\mathcal{L}^{\left(2\right)} & = & \frac{C_{5}C_{1}N}{1-2C_{5}C_{2}N} \hat{K}_{ij}\hat{K}^{ij}+\frac{1}{3}\frac{C_{1}N}{1+C_{2}N}K^{2}+\left(C_{3}+\frac{C_{4}}{N}\right)R. \label{calL_2_cub}
	\end{eqnarray}
We conclude that the Lagrangian (\ref{calL_2_cub}) propagates no scalar degree of freedom up to the second order in perturbations on a cosmological background.

\section{Conclusion} \label{sec:con}

In this work, we revisited the problem of propagating at most two tensorial degrees of freedom in a large class of spatially covariant gravity theories, of which the Lagrangian are polynomials built of spatial geometric quantities.
Although the general conditions have been derived in \cite{Gao:2019twq,Lin:2020nro}, these conditions are mathematically involved to be solved to yield concrete Lagrangians.

We thus take an alternative and complimentary approach in this work based on a perturbative analysis.
The idea is simple: if the Lagrangian has no scalar degree of freedom in a fully nonlinear sense, the scalar mode must not show up at any finite order if we perturbatively expand the Lagrangian around a cosmological background.
This perturbative analysis allows us to determine the coefficients in the Lagrangian order by order.
Since at the fully nonlinear level, there are finite number of conditions imposed on the functional form of the Lagrangian, this perturbative analysis must stop at some \emph{finite} order.
In other words, there must be a finite order up to which we ``kill'' the scalar mode, and then the scalar mode is eliminated at fully nonlinear order.
In fact, as being shown in \cite{Gao:2019lpz} in a specific example, it is sufficient to tune the coefficients up to the cubic order in the Lagrangian such that the unwanted scalar mode is fully removed.

In this work, we mainly focus on the linear cosmological perturbations.
In Sec. \ref{sec:pert_degen} we shown that in order to eliminate the unwanted scalar mode at the linear order, the degeneracy condition (\ref{degen_con}) must be imposed.
This is also supported by a more rigorous Lagrangian constraint analysis in Appendix \ref{app:clsmech}.
We then use (\ref{degen_con}) as the starting point to determine the coefficients of the Lagrangians for $d=2,3,4$ in Sec. \ref{sec:quad}, where $d$ is the total number of derivatives in a SCG monomial.
In particular, we have determined the concrete form of the Lagrangians for $d=2$ in (\ref{calL_2_fin}) and for $d=3$ in (\ref{calL_3_fin_1}) in the absence of $\nabla a$ terms and in (\ref{calL_3_fin_2}) in the presence of $\nabla a$ terms, respectively.
We thus conclude that (\ref{calL_2_fin}), (\ref{calL_3_fin_1}) and (\ref{calL_3_fin_2}) propagate no scalar mode at the linear order in perturbations around the cosmological background.
The scalar mode will re-arise in the naive combination of the scalar-mode-free Lagrangians for $d=2$ and $d=3$. 
Therefore one needs more restrictions on the coefficients in order to eliminate the scalar mode, if $d=2$ and $d=3$ Lagrangians are present simultaneously. 
The final results are given in (\ref{calL_23_fin_1}) in the absence of $\nabla a$ terms and in (\ref{calL_23_fin_2}) in the presence of $\nabla a$ terms, respectively.

It is not surprising that although the scalar mode has been eliminated at linear order, it may re-appear at nonlinear orders.
In Sec. \ref{sec:cub} we expand the Lagrangian up to the cubic order for $d=2$ and find the conditions for the coefficients to eliminate the scalar mode up to the cubic order.
The result is given in (\ref{calL_2_cub}).
In principle this procedure can be performed order by order, and one expects to determine the Lagrangian at some \emph{finite} order, such that the scalar mode has been fully eliminated.

\acknowledgments

This work was partly supported by the National Natural Science Foundation of China (NSFC) under the grant No. 11975020.

\appendix

\section{Classical mechanics with one dynamical and two auxiliary variables} \label{app:clsmech}

In this appendix, we make a thorough analysis of a classical mechanics system with one dynamical and two auxiliary variables.
We shall classify various cases according to the number of DOF's as well as the nature of constraints and gauge identities.

The most general quadratic Lagrangian for three variables $\left\{ q^{1},q^{2},q^{3}\right\} $,
of which one is dynamical and two are auxiliary variables, takes the
form
\begin{eqnarray}
	L & = & \frac{1}{2}g_{11}\left(\dot{q}^{1}\right)^{2}+f_{12}\dot{q}^{1}q^{2}+f_{13}\dot{q}^{1}q^{3}+\frac{1}{2}w_{22}\left(q^{2}\right)^{2}+\frac{1}{2}w_{33}\left(q^{3}\right)^{2}+w_{23}q^{2}q^{3}\nonumber \\
	&  & +\frac{1}{2}w_{11}\left(q^{1}\right)^{2}+w_{12}q^{1}q^{2}+w_{13}q^{1}q^{3}.\label{Lcons_1d2a_Lgen}
\end{eqnarray}
The coefficients $g_{11},f_{12}$ etc. are assumed to be constants
for simplicity. We assume $g_{11}\neq0$ so that $q^{1}$ acquires
an apparent kinetic term. While $q^{2}$ and $q^{3}$ do not have
explicit time derivatives and act as the auxiliary variables. Our task
is to search for cases in which there is no dynamics in the Lagrangian
(\ref{Lcons_1d2a_Lgen}). 

Varying the Lagrangian yields
\begin{equation}
	\delta L\simeq-\mathcal{E}_{i}^{(0)}\delta q^{i},
\end{equation}
where the equations of motion take the form
\begin{equation}
	\mathcal{E}_{i}^{(0)}\coloneqq W_{ij}^{(0)}\ddot{q}^{j}+V_{j}^{(0)}\approx0,
\end{equation}
with
\begin{equation}
	W_{ij}^{(0)}=\left(\begin{array}{ccc}
		g_{11} & 0 & 0\\
		0 & 0 & 0\\
		0 & 0 & 0
	\end{array}\right),
\end{equation}
and
\begin{eqnarray}
	V_{j}^{(0)} & = & \left(\begin{array}{c}
		f_{12}\dot{q}^{2}+f_{13}\dot{q}^{3}-w_{11}q^{1}-w_{12}q^{2}-w_{13}q^{3}\\
		-f_{12}\dot{q}^{1}-w_{12}q^{1}-w_{22}q^{2}-w_{23}q^{3}\\
		-f_{13}\dot{q}^{1}-w_{13}q^{1}-w_{23}q^{2}-w_{33}q^{3}
	\end{array}\right).
\end{eqnarray}
In this appendix, for clarity we use ``$\approx$'' to denote on-shell
equalities, i.e., those hold only when the equations of motion are
satisfied. Here and in what follows the superscript ``$^{(n)}$''
stands for ``level-$n$'', of which the meaning will be clear soon.

\subsection{Level-0}

Since $\mathrm{rank}\left(W_{ij}^{(0)}\right)=1$, there are 2 null-eigenvectors
for $W_{ij}^{(0)}$:
\begin{equation}
	u_{1,i}^{(0)}=\left(\begin{array}{c}
		0\\
		1\\
		0
	\end{array}\right),\qquad u_{2,i}^{(0)}=\left(\begin{array}{c}
		0\\
		0\\
		1
	\end{array}\right),\label{Lcons_1d2a_L0_u01u02}
\end{equation}
contracting which with $\mathcal{E}_{i}^{(0)}$ yields
\begin{eqnarray}
	u_{1}^{(0)i}\mathcal{E}_{i}^{(0)} & = & -f_{12}\dot{q}^{1}-w_{12}q^{1}-w_{22}q^{2}-w_{23}q^{3}\equiv\mathcal{E}_{2}^{(0)},\\
	u_{2}^{(0)i}\mathcal{E}_{i}^{(0)} & = & -f_{13}\dot{q}^{1}-w_{13}q^{1}-w_{23}q^{2}-w_{33}q^{3}\equiv\mathcal{E}_{3}^{(0)}.
\end{eqnarray}

According to the algorithm of detecting constraints in the Lagrangian
formalism, at each level, we have to examine whether the contractions
lead to constraints or identities. We have 3 cases according to how
many constraints/identities we get.

\subsubsection{Case 1: two identities}

If both contractions are vanishing identically, we get two gauge identities
$G_{1}^{(0)}\coloneqq\mathcal{E}_{2}^{(0)}\equiv0$ and $G_{2}^{(0)}\coloneqq\mathcal{E}_{3}^{(0)}\equiv0$
at ``level-0''. In this appendix ``$\equiv$'' stands for off-shell
identities, i.e., which always hold no matter the equations of motion
are satisfied or not. This requires $f_{12}=f_{13}=w_{12}=w_{13}=w_{22}=w_{23}=w_{33}=0$.
This case, however, is trivial since terms invovling $q^{2},q^{3}$
in the Lagrangian (\ref{Lcons_1d2a_Lgen}) completely drop out and
the Lagrangian reduces to that of a single variable $q^{1}$. Then
the algorithm ends. We include this case merely for completeness.

\subsubsection{Case 2: one constraint and one identity}

Without loss of generality, we assume at least one of $\left\{ f_{12},w_{12},w_{22},w_{23}\right\} $
is not vanishing, and denote the constraint at ``level-0'' as
\begin{equation}
	\phi^{(0)}\coloneqq u_{1}^{(0)i}\mathcal{E}_{i}^{(0)}\approx0.\label{Lcons_1d2a_c2_phi0}
\end{equation}

Then that $u_{2}^{(0)i}\mathcal{E}_{i}^{(0)}$ leads to an identity
implies that
\begin{equation}
	u_{2}^{(0)i}\mathcal{E}_{i}^{(0)}=\lambda\,u_{1}^{(0)i}\mathcal{E}_{i}^{(0)},\label{Lagcons_1d2a_u0E0_id}
\end{equation}
with some constant $\lambda$, i.e.,
\begin{eqnarray}
	f_{13} & = & \lambda\,f_{12},\label{Lcons_1d2a_L0_c2_1}\\
	w_{13} & = & \lambda\,w_{12},\label{Lcons_1d2a_L0_c2_2}\\
	w_{23} & = & \lambda\,w_{22},\label{Lcons_1d2a_L0_c2_3}\\
	w_{33} & = & \lambda\,w_{23}\equiv\lambda^{2}w_{22}.\label{Lcons_1d2a_L0_c2_4}
\end{eqnarray}
We then get one gauge identity at ``level-0'':
\begin{equation}
	G^{(0)}\coloneqq\lambda\,\mathcal{E}_{2}^{(0)}-\mathcal{E}_{3}^{(0)}\equiv0.\label{Lcons_1d2v_c2_G0}
\end{equation}
Note $\lambda=0$ is trivial, since in this case $q^{3}$-sector completely
drops out, and the original Lagrangian reduces to that of two variables
$q^{1}$ and $q^{2}$. 

\subsubsection{Case 3: two constraints}

This is of course the general case. As long as at least one of (\ref{Lcons_1d2a_L0_c2_1})-(\ref{Lcons_1d2a_L0_c2_4})
is not satisfied, we get two constraints at ``level-0'':
\begin{eqnarray}
	\phi_{1}^{(0)} & \coloneqq & \mathcal{E}_{2}^{(0)}\approx0,\label{Lcons_1d2v_c3_phi1}\\
	\phi_{2}^{(0)} & \coloneqq & \mathcal{E}_{3}^{(0)}\approx0.\label{Lcons_1d2v_c3_phi2}
\end{eqnarray}

\subsection{Case 2: level-1 \label{subsec:Lcons_1d2v_L1_c2}}

Using (\ref{Lcons_1d2a_L0_c2_1})-(\ref{Lcons_1d2a_L0_c2_4}) to replace
$\left\{ f_{13},w_{13},w_{23},w_{33}\right\} $ in terms of $\left\{ f_{12},w_{12},w_{22},w_{23}\right\} $,
the constraint $\phi^{(0)}$ in (\ref{Lcons_1d2a_c2_phi0}) becomes
\begin{equation}
	\phi^{(0)}\rightarrow-f_{12}\dot{q}^{1}-w_{12}q^{1}-w_{22}q^{2}-\lambda w_{22}q^{3}.\label{Lcons_1d2a_c2_phi0_fin}
\end{equation}
According the the standard algorithm, we build the enlarged ``vector''
of equations of motion as
\begin{equation}
	\mathcal{E}_{i_{1}}^{(1)}\coloneqq\left(\begin{array}{c}
		\mathcal{E}_{i}^{(0)}\\
		\dot{\phi}^{(0)}
	\end{array}\right)=W_{i_{1}j}^{(1)}\ddot{q}^{j}+V_{i_{1}}^{(1)},
\end{equation}
with
\begin{equation}
	W_{i_{1}j}^{(1)}=\left(\begin{array}{ccc}
		g_{11} & 0 & 0\\
		0 & 0 & 0\\
		0 & 0 & 0\\
		-f_{12} & 0 & 0
	\end{array}\right),
\end{equation}
and
\begin{equation}
	V_{i_{1}}^{(1)}=\left(\begin{array}{c}
		f_{12}\dot{q}^{2}+\lambda f_{12}\dot{q}^{3}-w_{11}q^{1}-w_{12}q^{2}-\lambda w_{12}q^{3}\\
		-f_{12}\dot{q}^{1}-w_{12}q^{1}-w_{22}q^{2}-\lambda w_{22}q^{3}\\
		-\lambda f_{12}\dot{q}^{1}-\lambda w_{12}q^{1}-\lambda w_{22}q^{2}-\lambda^{2}w_{22}q^{3}\\
		-w_{12}\dot{q}^{1}-w_{22}\dot{q}^{2}-\lambda w_{22}\dot{q}^{3}
	\end{array}\right).
\end{equation}

There are two trivial null-eigenvectors for $W_{i_{1}j}^{(1)}$, which
are merely $u_{1,i}^{(0)},u_{2,i}^{(0)}$ in (\ref{Lcons_1d2a_L0_u01u02})
augmented by zero's. On the other hand, there is a non-trivial null-eigenvector
\begin{equation}
	u_{i_{1}}^{(1)}=\left(\begin{array}{c}
		f_{12}\\
		0\\
		0\\
		g_{11}
	\end{array}\right),
\end{equation}
which is valid no matter $f_{12}=0$ or not. Contracting $u_{i_{1}}^{(1)}$
with $\mathcal{E}_{i_{1}}^{(1)}$ yields
\begin{eqnarray}
	u^{(1)i_{1}}\mathcal{E}_{i_{1}}^{(1)} & = & -g_{11}w_{12}\dot{q}^{1}+\left(f_{12}^{2}-g_{11}w_{22}\right)\dot{q}^{2}+\lambda\left(f_{12}^{2}-g_{11}w_{22}\right)\dot{q}^{3}\nonumber \\
	&  & -f_{12}w_{11}q^{1}-f_{12}w_{12}q^{2}-\lambda f_{12}w_{12}q^{3}.
\end{eqnarray}
We need to check whether $u^{(1)i_{1}}\mathcal{E}_{i_{1}}^{(1)}$
leads to an identity or a new constriant. There are two sub-cases.

\subsubsection{Case 2.1: one identity}

Since up to ``level-1'', we have only one constraint $\phi^{(0)}$
given in (\ref{Lcons_1d2a_c2_phi0_fin}). If $u^{(1)i_{1}}\mathcal{E}_{i_{1}}^{(1)}$
is not an independent constraint, it implies that
\begin{equation}
	u^{(1)i_{1}}\mathcal{E}_{i_{1}}^{(1)}\propto\phi_{1}^{(0)},\label{Lagcons_1d2a_u1E1_id}
\end{equation}
which puts restrictions on the coefficients. After some manipulation,
the necessary and sufficient condition for (\ref{Lagcons_1d2a_u1E1_id})
can be written as 
\begin{equation}
	w_{12}=\sqrt{w_{11}}\,\frac{f_{12}}{\sqrt{g_{11}}},\qquad w_{22}=\frac{f_{12}^{2}}{g_{11}},\label{Lcons_1d2a_c2.1_w_fin}
\end{equation}
with $f_{12}\neq0$. Note we must have $f_{12}\neq0$ since if $f_{12}=0$,
(\ref{Lagcons_1d2a_u1E1_id}) implies $w_{12}=w_{22}=w_{23}=0$, which
is conflict with the assumption that at least one of $\left\{ f_{12},w_{12},w_{22},w_{23}\right\} $
is not vanishing in order to have the constraint $\phi^{(0)}$. Then
we get one gauge identity at ``level-1'':
\begin{eqnarray}
	G^{(1)} & := & u^{(1)i_{1}}\mathcal{E}_{i_{1}}^{(1)}-g_{11}w_{11}\,\phi_{1}^{(0)}\nonumber \\
	& = & f_{12}\mathcal{E}_{1}^{(0)}+g_{11}\frac{\mathrm{d}\mathcal{E}_{2}^{(0)}}{\mathrm{d}t}-g_{11}w_{11}\,\mathcal{E}_{2}^{(0)}\equiv0.\label{Lcons_1d2v_G1-1}
\end{eqnarray}
Then the algorithm ends. 

In this case, we have one constraint $\phi_{1}^{(0)}$ in (\ref{Lcons_1d2a_c2_phi0}),
two gauge identities $G^{(0)}$ and $G^{(1)}$ in (\ref{Lcons_1d2v_c2_G0})
and (\ref{Lcons_1d2v_G1-1}), respectively. It is easy to show that
in this case there is no dynamical degree of freedom.

\subsubsection{Case 2.2: one constraint}

As long as (\ref{Lcons_1d2a_c2.1_w_fin}) (at least one of the two
equalities) is not\textcolor{red}{{} }satisfied, $u^{(1)i_{1}}\mathcal{E}_{i_{1}}^{(1)}$
leads to a new independent constraint:
\begin{equation}
	\phi^{(1)}:=u^{(1)i_{1}}\mathcal{E}_{i_{1}}^{(1)}\approx0.\label{Lcons_1d2v_c2.2_phi1}
\end{equation}
Then we go to the next level.

\subsection{Case 2.2: level-2 \label{subsec:Lcons_1d2v_L2_c2.2}}

By appending $\dot{\phi}^{(1)}$ (with $\phi^{(1)}$ in (\ref{Lcons_1d2v_c2.2_phi1}))
to $\mathcal{E}_{i_{1}}^{(1)}$, we build the enlarged vector of equations
of motion: 
\begin{equation}
	\mathcal{E}_{i_{2}}^{(2)}=W_{i_{2}j}^{(2)}\ddot{q}^{j}+V_{i_{2}}^{(2)},
\end{equation}
with
\begin{equation}
	W_{i_{2}j}^{(2)}=\left(\begin{array}{ccc}
		g_{11} & 0 & 0\\
		0 & 0 & 0\\
		0 & 0 & 0\\
		-f_{12} & 0 & 0\\
		-g_{11}w_{12} & f_{12}^{2}-g_{11}w_{22} & \lambda\left(f_{12}^{2}-g_{11}w_{22}\right)
	\end{array}\right),
\end{equation}
and
\begin{equation}
	V_{i_{2}}^{(2)}=\left(\begin{array}{c}
		f_{12}\dot{q}^{2}+\lambda f_{12}\dot{q}^{3}-w_{11}q^{1}-w_{12}q^{2}-\lambda w_{12}q^{3}\\
		-f_{12}\dot{q}^{1}-w_{12}q^{1}-w_{22}q^{2}-\lambda w_{22}q^{3}\\
		-\lambda f_{12}\dot{q}^{1}-\lambda w_{12}q^{1}-\lambda w_{22}q^{2}-\lambda^{2}w_{22}q^{3}\\
		-w_{12}\dot{q}^{1}-w_{22}\dot{q}^{2}-\lambda w_{22}\dot{q}^{3}\\
		-f_{12}w_{11}\dot{q}^{1}-f_{12}w_{12}\dot{q}^{2}-\lambda f_{12}w_{12}\dot{q}^{3}
	\end{array}\right).
\end{equation}
Since we have assumed $\lambda\neq0$, then we have two sub-cases.

\subsubsection{Case 2.2.1: $w_{22}\protect\neq\frac{f_{12}^{2}}{g_{11}}$}

In this case, $W_{i_{2}j}^{(2)}$ do not possesses further non-trivial
left null-eigenvector. The algorithm therefore ends. 

In this case, we have two constraints $\phi^{(0)},\phi^{(1)}$ and
one gauge identity $G^{(0)}$. One can show that there is one dynamical
degree of freedom.

\subsubsection{Case 2.2.2: $w_{22}=\frac{f_{12}^{2}}{g_{11}}$ while $w_{12}\protect\neq\sqrt{w_{11}}\,\frac{f_{12}}{\sqrt{g_{11}}}$}

Since $w_{22}=\frac{f_{12}^{2}}{g_{11}}$, we have to require that
the other equality in (\ref{Lcons_1d2a_c2.1_w_fin}) is not satisfied,
i.e.,
\begin{equation}
	w_{12}\neq\sqrt{w_{11}}\,\frac{f_{12}}{\sqrt{g_{11}}}.\label{Lcons_1d2a_c2.2.2_w}
\end{equation}
In this case, $W_{i_{2}j}^{(2)}$ possesses a new non-trivial null-eigenvector
\begin{equation}
	u_{i_{2}}^{(2)}=\left(\begin{array}{c}
		w_{12}\\
		0\\
		0\\
		0\\
		1
	\end{array}\right).
\end{equation}
Contracting $u_{i_{2}}^{(2)}$ with $\mathcal{E}_{i_{2}}^{(2)}$ yields
\begin{equation}
	u^{(2)i_{2}}\mathcal{E}_{i_{2}}^{(2)}=-f_{12}w_{11}\dot{q}^{1}-w_{12}w_{11}q^{1}-w_{12}^{2}q^{2}-\lambda w_{12}^{2}q^{3}.
\end{equation}

Recall that we have two constraints $\phi_{1}^{(0)}$ in (\ref{Lcons_1d2a_c2_phi0_fin})
and $\phi^{(1)}$ in (\ref{Lcons_1d2v_c2.2_phi1}), which in our case
reduce to be
\begin{eqnarray}
	\phi^{(0)} & \rightarrow & -f_{12}\dot{q}^{1}-w_{12}q^{1}-\frac{f_{12}^{2}}{g_{11}}q^{2}-\lambda\,\frac{f_{12}^{2}}{g_{11}}q^{3},\label{Lcons_1d2a_c2.2.2_phi0}\\
	\phi^{(1)} & \rightarrow & -g_{11}w_{12}\dot{q}^{1}-f_{12}w_{11}q^{1}-f_{12}w_{12}q^{2}-\lambda f_{12}w_{12}q^{3}.\label{Lcons_1d2a_c2.2.2_phi1}
\end{eqnarray}
One can show that $u^{(2)i_{2}}\mathcal{E}_{i_{2}}^{(2)}$ is linearly
independent of $\phi^{(0)}$ and $\phi^{(1)}$ as long as (\ref{Lcons_1d2a_c2.2.2_w})
is satisfied. Therefore $u^{(2)i_{2}}\mathcal{E}_{i_{2}}^{(2)}$ leads
to a new constraint
\begin{equation}
	\phi^{(2)}:=u^{(2)i_{2}}\mathcal{E}_{i_{2}}^{(2)}\approx0.\label{Lcons_1d2a_c2.2.2_phi2}
\end{equation}

\subsection{Case 2.2.2: level-3 \label{subsec:Lcons_1d2v_L3_c2.2.2}}

Appending $\dot{\phi}^{(2)}$ (with $\phi^{(2)}$ given in (\ref{Lcons_1d2a_c2.2.2_phi2}))
to $\mathcal{E}_{i_{2}}^{(2)}$ yields
\begin{equation}
	\mathcal{E}_{i_{3}}^{(3)}=W_{i_{3}j}^{(3)}\ddot{q}^{j}+V_{i_{3}}^{(3)},
\end{equation}
with
\begin{equation}
	W_{i_{3}j}^{(3)}=\left(\begin{array}{ccc}
		g_{11} & 0 & 0\\
		0 & 0 & 0\\
		0 & 0 & 0\\
		-f_{12} & 0 & 0\\
		-g_{11}w_{12} & 0 & 0\\
		-f_{12}w_{11} & 0 & 0
	\end{array}\right),
\end{equation}
and
\begin{equation}
	V_{i_{3}}^{(3)}=\left(\begin{array}{c}
		f_{12}\dot{q}^{2}+\lambda f_{12}\dot{q}^{3}-w_{11}q^{1}-w_{12}q^{2}-\lambda w_{12}q^{3}\\
		-f_{12}\dot{q}^{1}-w_{12}q^{1}-\frac{f_{12}^{2}}{g_{11}}q^{2}-\lambda\frac{f_{12}^{2}}{g_{11}}q^{3}\\
		-\lambda f_{12}\dot{q}^{1}-\lambda w_{12}q^{1}-\lambda\frac{f_{12}^{2}}{g_{11}}q^{2}-\lambda^{2}\frac{f_{12}^{2}}{g_{11}}q^{3}\\
		-w_{12}\dot{q}^{1}-\frac{f_{12}^{2}}{g_{11}}\dot{q}^{2}-\lambda\frac{f_{12}^{2}}{g_{11}}\dot{q}^{3}\\
		-f_{12}w_{11}\dot{q}^{1}-f_{12}w_{12}\dot{q}^{2}-\lambda f_{12}w_{12}\dot{q}^{3}\\
		-w_{11}w_{12}\dot{q}^{1}-w_{12}^{2}\dot{q}^{2}-\lambda w_{12}^{2}\dot{q}^{3}
	\end{array}\right).
\end{equation}

$W_{i_{3}j}^{(3)}$ possesses a non-trivial null-eigenvector
\begin{equation}
	u_{i_{3}}^{(3)}=\left(\begin{array}{c}
		f_{12}w_{11}\\
		0\\
		0\\
		0\\
		0\\
		g_{11}
	\end{array}\right),\label{Lcons_1d2a_L3_c2.2.2_u3}
\end{equation}
contracting which with $\mathcal{E}_{i_{3}}^{(3)}$ yields
\begin{eqnarray}
	u^{(3)i_{3}}\mathcal{E}_{i_{3}}^{(3)} & = & -g_{11}w_{11}w_{12}\dot{q}^{1}+\left(w_{11}f_{12}^{2}-g_{11}w_{12}^{2}\right)\dot{q}^{2}+\lambda\left(w_{11}f_{12}^{2}-g_{11}w_{12}^{2}\right)\dot{q}^{3}\nonumber \\
	&  & -f_{12}w_{11}^{2}q^{1}-f_{12}w_{11}w_{12}q^{2}-\lambda f_{12}w_{11}w_{12}q^{3}.
\end{eqnarray}
Comparing with the previous constaints $\phi^{(0)},\phi^{(1)},\phi^{(2)}$
in (\ref{Lcons_1d2a_c2_phi0}), (\ref{Lcons_1d2v_c2.2_phi1}) and
(\ref{Lcons_1d2a_c2.2.2_phi2}), clearly $u^{(3)i_{3}}\mathcal{E}_{i_{3}}^{(3)}$
leads to a new constraint
\begin{equation}
	\phi^{(3)}:=u^{(3)i_{3}}\mathcal{E}_{i_{3}}^{(3)}\approx0,\label{Lcons_1d2a_c2.2.2_phi3}
\end{equation}
since\textcolor{red}{{} ${\normalcolor w_{11}f_{12}^{2}-g_{11}w_{12}^{2}\neq0}$}.

\subsection{Case 2.2.2: level-4 \label{subsec:Lcons_1d2v_L4_c2.2.2}}

By appending $\dot{\phi}^{(3)}$ (with $\phi^{(3)}$ given in (\ref{Lcons_1d2a_c2.2.2_phi3}))
to $\mathcal{E}_{i_{3}}^{(3)}$ yields
\begin{equation}
	\mathcal{E}_{i_{4}}^{(4)}=W_{i_{4}j}^{(4)}\ddot{q}^{j}+V_{i_{4}}^{(4)},
\end{equation}
with
\begin{equation}
	W_{i_{4}j}^{(4)}=\left(\begin{array}{ccc}
		g_{11} & 0 & 0\\
		0 & 0 & 0\\
		0 & 0 & 0\\
		-f_{12} & 0 & 0\\
		-g_{11}w_{12} & 0 & 0\\
		-f_{12}w_{11} & 0 & 0\\
		-g_{11}w_{12}w_{11} & w_{11}f_{12}^{2}-g_{11}w_{12}^{2} & \lambda\left(w_{11}f_{12}^{2}-g_{11}w_{12}^{2}\right)
	\end{array}\right),
\end{equation}
and
\begin{equation}
	V_{i_{4}}^{(4)}=\left(\begin{array}{c}
		f_{12}\dot{q}^{2}+\lambda f_{12}\dot{q}^{3}-w_{11}q^{1}-w_{12}q^{2}-\lambda w_{12}q^{3}\\
		-f_{12}\dot{q}^{1}-w_{12}q^{1}-\frac{f_{12}^{2}}{g_{11}}q^{2}-\lambda\frac{f_{12}^{2}}{g_{11}}q^{3}\\
		-\lambda f_{12}\dot{q}^{1}-\lambda w_{12}q^{1}-\lambda\frac{f_{12}^{2}}{g_{11}}q^{2}-\lambda^{2}\frac{f_{12}^{2}}{g_{11}}q^{3}\\
		-w_{12}\dot{q}^{1}-\frac{f_{12}^{2}}{g_{11}}\dot{q}^{2}-\lambda\frac{f_{12}^{2}}{g_{11}}\dot{q}^{3}\\
		-f_{12}w_{11}\dot{q}^{1}-f_{12}w_{12}\dot{q}^{2}-\lambda f_{12}w_{12}\dot{q}^{3}\\
		-w_{11}w_{12}\dot{q}^{1}-w_{12}^{2}\dot{q}^{2}-\lambda w_{12}^{2}\dot{q}^{3}\\
		-f_{12}w_{11}^{2}\dot{q}^{1}-f_{12}w_{11}w_{12}\dot{q}^{2}-\lambda f_{12}w_{11}w_{12}\dot{q}^{3}
	\end{array}\right).
\end{equation}
Clearly, since $w_{11}f_{12}^{2}-g_{11}w_{12}^{2}\neq0$, $W_{i_{4}j}^{(4)}$
possesses no non-trivial null-eigenvector. The algorithm ends. 

To summarize, in this case, we have four constraints: $\phi^{(0)},\phi^{(1)},\phi^{(2)},\phi^{(3)}$
given in (\ref{Lcons_1d2a_c2.2.2_phi0}), (\ref{Lcons_1d2a_c2.2.2_phi1}),
(\ref{Lcons_1d2a_c2.2.2_phi2}) and (\ref{Lcons_1d2a_c2.2.2_phi3}),
and one gauge identity $G^{(0)}$ given in (\ref{Lcons_1d2v_c2_G0}),
respectively. As a result, there is no dynamical degree of freedom.

\subsection{Case 3: level-1 \label{subsec:Lcons_1d2v_L1_c3}}

From (\ref{Lcons_1d2v_c3_phi1}) and (\ref{Lcons_1d2v_c3_phi2}),
the two independent constaints at ``level-0'' are:
\begin{eqnarray}
	\dot{\phi}_{1}^{(0)} & = & -f_{12}\ddot{q}^{1}-w_{12}\dot{q}^{1}-w_{22}\dot{q}^{2}-w_{23}\dot{q}^{3},\\
	\dot{\phi}_{2}^{(0)} & = & -f_{13}\ddot{q}^{1}-w_{13}\dot{q}^{1}-w_{23}\dot{q}^{2}-w_{33}\dot{q}^{3}.
\end{eqnarray}

In the ``case-3'', the enlarged vector of equations of motion is
\begin{equation}
	\mathcal{E}_{i_{1}}^{(1)}=\left(\begin{array}{c}
		\mathcal{E}_{i}^{(0)}\\
		\dot{\phi}_{1}^{(0)}\\
		\dot{\phi}_{2}^{(0)}
	\end{array}\right)=W_{i_{1}j}^{(1)}\ddot{q}^{j}+V_{i_{1}}^{(1)},
\end{equation}
with
\begin{equation}
	W_{i_{1}j}^{(1)}=\left(\begin{array}{ccc}
		g_{11} & 0 & 0\\
		0 & 0 & 0\\
		0 & 0 & 0\\
		-f_{12} & 0 & 0\\
		-f_{13} & 0 & 0
	\end{array}\right),
\end{equation}
and
\begin{equation}
	V_{i_{1}}^{(1)}=\left(\begin{array}{c}
		f_{12}\dot{q}^{2}+f_{13}\dot{q}^{3}-w_{11}q^{1}-w_{12}q^{2}-w_{13}q^{3}\\
		-f_{12}\dot{q}^{1}-w_{12}q^{1}-w_{22}q^{2}-w_{23}q^{3}\\
		-f_{13}\dot{q}^{1}-w_{13}q^{1}-w_{23}q^{2}-w_{33}q^{3}\\
		-w_{12}\dot{q}^{1}-w_{22}\dot{q}^{2}-w_{23}\dot{q}^{3}\\
		-w_{13}\dot{q}^{1}-w_{23}\dot{q}^{2}-w_{33}\dot{q}^{3}
	\end{array}\right).
\end{equation}
In this case $W_{i_{1}j}^{(1)}$ possesses 2 non-trivial null-eigenvectors:
\begin{equation}
	u_{1,i_{1}}^{(1)}=\left(\begin{array}{c}
		f_{12}\\
		0\\
		0\\
		g_{11}\\
		0
	\end{array}\right),\qquad u_{2,i_{1}}^{(1)}=\left(\begin{array}{c}
		f_{13}\\
		0\\
		0\\
		0\\
		g_{11}
	\end{array}\right).
\end{equation}
Contracting $u_{1,i_{1}}^{(1)}$ and $u_{2,i_{1}}^{(1)}$ with $\mathcal{E}_{i_{1}}^{(1)}$
yields
\begin{eqnarray}
	u_{1}^{(1)i_{1}}\mathcal{E}_{i_{1}}^{(1)} & = & -g_{11}w_{12}\dot{q}^{1}+\left(f_{12}^{2}-g_{11}w_{22}\right)\dot{q}^{2}+\left(f_{12}f_{13}-g_{11}w_{23}\right)\dot{q}^{3}\nonumber \\
	&  & -f_{12}w_{11}q^{1}-f_{12}w_{12}q^{2}-f_{12}w_{13}q^{3},
\end{eqnarray}
and
\begin{eqnarray}
	u_{2}^{(1)i_{1}}\mathcal{E}_{i_{1}}^{(1)} & = & -g_{11}w_{13}\dot{q}^{1}+\left(f_{13}f_{12}-g_{11}w_{23}\right)\dot{q}^{2}+\left(f_{13}^{2}-g_{11}w_{33}\right)\dot{q}^{3}\nonumber \\
	&  & -f_{13}w_{11}q^{1}-f_{13}w_{12}q^{2}-f_{13}w_{13}q^{3}.
\end{eqnarray}

We need to check whether $u_{1}^{(1)i_{1}}\mathcal{E}_{i_{1}}^{(1)}$
and $u_{2}^{(1)i_{1}}\mathcal{E}_{i_{1}}^{(1)}$ lead to new constraints
or identities. To this end, together with $\phi_{1}^{(0)}$ and $\phi_{2}^{(0)}$
in (\ref{Lcons_1d2v_c3_phi1}) and (\ref{Lcons_1d2v_c3_phi2}), we
write
\begin{equation}
	\left(\begin{array}{c}
		\phi_{1}^{(0)}\\
		\phi_{2}^{(0)}\\
		u_{1}^{(1)i_{1}}\mathcal{E}_{i_{1}}^{(1)}\\
		u_{2}^{(1)i_{1}}\mathcal{E}_{i_{1}}^{(1)}
	\end{array}\right)=\bm{M}^{(1)}\left(\begin{array}{c}
		\dot{q}^{1}\\
		\dot{q}^{2}\\
		\dot{q}^{3}\\
		q^{1}\\
		q^{2}\\
		q^{3}
	\end{array}\right),
\end{equation}
with the $4\times6$ matrix
\begin{equation}
	\bm{M}^{(1)}:=\left(\begin{array}{cccccc}
		-f_{12} & 0 & 0 & -w_{12} & -w_{22} & -w_{23}\\
		-f_{13} & 0 & 0 & -w_{13} & -w_{23} & -w_{33}\\
		-g_{11}w_{12} & f_{12}^{2}-g_{11}w_{22} & f_{12}f_{13}-g_{11}w_{23} & -f_{12}w_{11} & -f_{12}w_{12} & -f_{12}w_{13}\\
		-g_{11}w_{13} & f_{12}f_{13}-g_{11}w_{23} & f_{13}^{2}-g_{11}w_{33} & -f_{13}w_{11} & -f_{13}w_{12} & -f_{13}w_{13}
	\end{array}\right).
\end{equation}
Then the question is equivalent to checking the rank of $\bm{M}^{(1)}$.
For later convenience, we define the sub-matrix 
\begin{equation}
	\bm{\Delta}:=\left(\begin{array}{cc}
		f_{12}^{2}-g_{11}w_{22} & f_{12}f_{13}-g_{11}w_{23}\\
		f_{12}f_{13}-g_{11}w_{23} & f_{13}^{2}-g_{11}w_{33}
	\end{array}\right),\label{Lcons_L1_c3_Delta}
\end{equation}
of which the determinant is
\begin{equation}
	\det\bm{\Delta}=g_{11}\left[-f_{12}^{2}w_{33}+2f_{12}f_{13}w_{23}-f_{13}^{2}w_{22}+g_{11}\left(w_{22}w_{33}-w_{23}^{2}\right)\right].
\end{equation}

\subsubsection{2 identities (impossible) \label{par:Lcons_2d1a_L1_c3_2id}}

First we shall show that it is impossible to have 2 identities. In
fact, in order to have 2 new identities, we have to require that the
entries of the sub-matrix $\bm{\Delta}$ in (\ref{Lcons_L1_c3_Delta})
are vanishing identically. This can be also understood that since
there are neither $\dot{q}^{2}$ nor $\dot{q}^{3}$ terms in $\phi_{1}^{(0)},\phi_{2}^{(2)}$,
as long as at least one of the coefficients of $\dot{q}^{2}$ or $\dot{q}^{3}$
in $u_{1}^{(1)i_{1}}\mathcal{E}_{i_{1}}^{(1)}$ and $u_{2}^{(1)i_{1}}\mathcal{E}_{i_{1}}^{(1)}$
is not vanishing, we get a new constraint. To conclude, the necessary
conditions to have two new identities are
\begin{equation}
	w_{22}=\frac{f_{12}^{2}}{g_{11}},\qquad w_{23}=\frac{f_{12}f_{13}}{g_{11}},\qquad w_{33}=\frac{f_{13}^{2}}{g_{11}}.\label{Lcons_1d2v_2id_w22w23w33}
\end{equation}
On the other hand, comparing with (\ref{Lcons_1d2a_L0_c2_1})-(\ref{Lcons_1d2a_L0_c2_4}),
we have to require
\begin{equation}
	f_{12}w_{13}-f_{13}w_{12}\neq0,\label{Lcons_1d2v_2id_fw}
\end{equation}
in order not to go back to ``case 2''.

With these considerations, $u_{1}^{(1)i_{1}}\mathcal{E}_{i_{1}}^{(1)}$
and $u_{2}^{(1)i_{1}}\mathcal{E}_{i_{1}}^{(1)}$ reduce to be
\begin{eqnarray}
	u_{1}^{(1)i_{1}}\mathcal{E}_{i_{1}}^{(1)} & \rightarrow & -g_{11}w_{12}\dot{q}^{1}-f_{12}w_{11}q^{1}-f_{12}w_{12}q^{2}-f_{12}w_{13}q^{3},\\
	u_{2}^{(1)i_{1}}\mathcal{E}_{i_{1}}^{(1)} & \rightarrow & -g_{11}w_{13}\dot{q}^{1}-f_{13}w_{11}q^{1}-f_{13}w_{12}q^{2}-f_{13}w_{13}q^{3}.
\end{eqnarray}
In order to have two identities among the 4 equalities $\left\{ \phi_{1}^{(0)},\phi_{2}^{(0)},u_{1}^{(1)i_{1}}\mathcal{E}_{i_{1}}^{(1)},u_{2}^{(1)i_{1}}\mathcal{E}_{i_{1}}^{(1)}\right\} \approx0$
we have to make sure that the rank of the $4\times4$ matrix 
\begin{equation}
	\left(\begin{array}{cccc}
		-f_{12} & -w_{12} & -\frac{f_{12}^{2}}{g_{11}} & -\frac{f_{12}f_{13}}{g_{11}}\\
		-f_{13} & -w_{13} & -\frac{f_{12}f_{13}}{g_{11}} & -\frac{f_{13}^{2}}{g_{11}}\\
		-g_{11}w_{12} & -f_{12}w_{11} & -f_{12}w_{12} & -f_{12}w_{13}\\
		-g_{11}w_{13} & -f_{13}w_{11} & -f_{13}w_{12} & -f_{13}w_{13}
	\end{array}\right)
\end{equation}
is 2. However, the determinant of the above $4\times4$ matrix is
\begin{equation}
	-\left(f_{13}w_{12}-f_{12}w_{13}\right)^{3}\neq0,\label{Lcons_1d2a_L1_2id_detsub}
\end{equation}
since we must have (\ref{Lcons_1d2v_2id_fw}). Therefore it is impossible
to to have two independent constraints $\left\{ \phi_{1}^{(0)},\phi_{2}^{(0)}\right\} $
at ``level-0'', and in the meanwhile to have two new gauge identities
$\left\{ G_{1}^{(1)},G_{2}^{(1)}\right\} $ at ``level-1''.

\subsubsection{Case 3.1: one constraint and one identity \label{par:Lcons_1d1a_L1_c3.1}}

The necessary condition is that the sub-matrix $\bm{\Delta}$ defined
in (\ref{Lcons_L1_c3_Delta}) is degenerate but not identically vanishing,
i.e., 
\begin{equation}
	\mathrm{rank}\bm{\Delta}=1.
\end{equation}
This is because:
\begin{itemize}
	\item if the $\mathrm{rank}\bm{\Delta}=2$, there will be 2 new constraints. 
	\item if the $\mathrm{rank}\bm{\Delta}=0$ (then $\bm{\Delta}\equiv0$),
	according to the analysis in Sec. \ref{par:Lcons_2d1a_L1_c3_2id},
	either there are still two new constraints when (\ref{Lcons_1d2v_2id_fw})
	is satisfied, or we go back to ``case 2.1''.
\end{itemize}
Without loss of generality, the sub-matrix $\bm{\Delta}$ can be written
in the form
\begin{equation}
	\bm{\Delta}=\omega\left(\begin{array}{cc}
		1 & \eta\\
		\eta & \eta^{2}
	\end{array}\right),\label{Lcons_1d2a_L1_c3.1_gDform}
\end{equation}
with $\omega,\eta$ being constants. Note we require $\omega\neq0$
in order to have $\mathrm{rank}\bm{\Delta}=1$. With (\ref{Lcons_1d2a_L1_c3.1_gDform}),
$\bm{M}^{(1)}$ reduces to be
\begin{equation}
	\bm{M}^{(1)}\rightarrow\left(\begin{array}{cccccc}
		-f_{12} & 0 & 0 & -w_{12} & \frac{\omega-f_{12}^{2}}{g_{11}} & \frac{\eta\omega-f_{12}f_{13}}{g_{11}}\\
		-f_{13} & 0 & 0 & -w_{13} & \frac{\eta\omega-f_{12}f_{13}}{g_{11}} & \frac{\eta^{2}\omega-f_{13}^{2}}{g_{11}}\\
		-g_{11}w_{12} & \omega & \omega\eta & -f_{12}w_{11} & -f_{12}w_{12} & -f_{12}w_{13}\\
		-g_{11}w_{13} & \omega\eta & \omega\eta^{2} & -f_{13}w_{11} & -f_{13}w_{12} & -f_{13}w_{13}
	\end{array}\right),
\end{equation}
our question thus reduces to checking if it is possible to have $\mathrm{rank}\bm{M}^{(1)}=3$.

One can show that the necessary condition to have $\mathrm{rank}\bm{M}^{(1)}=3$
is to require
\begin{equation}
	\mathcal{D}^{(1)}\coloneqq-\omega\left(f_{13}w_{12}-f_{12}w_{13}\right)^{2}+\omega^{2}\left[\left(\eta w_{12}-w_{13}\right)^{2}-\frac{w_{11}}{g_{11}}\left(\eta f_{12}-f_{13}\right)^{2}\right]=0.\label{Lcons_1d2a_L1_c3.1_deg}
\end{equation}
If 
\begin{equation}
	f_{13}w_{12}-f_{12}w_{13}=0,\label{Lcons_1d2a_L1_c3.1_w13_cd}
\end{equation}
then we need to require
\begin{equation}
	f_{13}-\eta\,f_{12}\neq0,\label{Lcons_1d2a_L1_c3.1_f}
\end{equation}
and (\ref{Lcons_1d2a_L1_c3.1_deg}) yields 
\begin{equation}
	w_{12}=\frac{f_{12}\sqrt{w_{11}}}{\sqrt{g_{11}}}.\label{Lcons_1d2a_L1_c3.1_w12}
\end{equation}
On the other hand, if 
\begin{equation}
	f_{13}w_{12}-f_{12}w_{13}\neq0,
\end{equation}
(\ref{Lcons_1d2a_L1_c3.1_deg}) also implies one constraint among
the coefficients. In both cases we have $\mathrm{rank}\bm{M}^{(1)}=3$.

To conclude, it is possible to have $\mathrm{rank}\bm{M}^{(1)}=3$
so that we get one constraint and one gauge identity on ``level-1''.
Without loss of generality, we may choose the new constraint to be
\begin{equation}
	\phi^{(1)}:=u_{1}^{(1)i_{1}}\mathcal{E}_{i_{1}}^{(1)}\approx0.\label{Lcons_1d2a_c3.1_phi1}
\end{equation}
The gauge identity must be the form
\begin{equation}
	G^{(1)}:=a\,\phi_{1}^{(0)}+b\,\phi_{2}^{(1)}-\eta\,\phi_{1}^{(1)}+u_{2}^{(1)i_{1}}\mathcal{E}_{i_{1}}^{(1)},\label{Lcons_1d2a_c3.1_G1}
\end{equation}
where the constants $a,b$ are not vanishing simultaneously. $a$
and $b$ are determined by the concrete form of the null-eigenvector
of $\bm{M}^{(1)}$, which we do not show here explicitly.

\subsubsection{Case 3.2: two new constraints}

Generally, either 
\begin{enumerate}
	\item $\det\bm{\Delta}\neq0$, or 
	\item $\mathrm{rank}\bm{\Delta}=1$, $\mathcal{D}^{(1)}\neq0$ (with $\mathcal{D}^{(1)}$
	defined in (\ref{Lcons_1d2a_L1_c3.1_deg})), or
	\item $\bm{\Delta}=0$ and $f_{13}w_{12}-f_{12}w_{13}\neq0$,
\end{enumerate}
we have two constraints at ``level-1'':
\begin{equation}
	\phi_{1}^{(1)}:=u_{1}^{(1)i_{1}}\mathcal{E}_{i_{1}}^{(1)}\approx0,\label{Lcons_1d2v_c3.2_phi11}
\end{equation}
and
\begin{equation}
	\phi_{2}^{(1)}:=u_{2}^{(1)i_{1}}\mathcal{E}_{i_{1}}^{(1)}\approx0.\label{Lcons_1d2v_c3.2_phi12}
\end{equation}

\subsection{Case 3.1: level-2 \label{subsec:Lcons_1d2v_L2_c3.1}}

By appending $\dot{\phi}^{(1)}$ (with $\phi^{(1)}$ given in (\ref{Lcons_1d2a_c3.1_phi1}))
to $\mathcal{E}_{i_{1}}^{(1)}$ we have
\begin{equation}
	\mathcal{E}_{i_{2}}^{(2)}=W_{i_{2}j}^{(2)}\ddot{q}^{j}+V_{i_{2}}^{(2)},
\end{equation}
with
\begin{equation}
	W_{i_{2}j}^{(2)}=\left(\begin{array}{ccc}
		g_{11} & 0 & 0\\
		0 & 0 & 0\\
		0 & 0 & 0\\
		-f_{12} & 0 & 0\\
		-f_{13} & 0 & 0\\
		-g_{11}w_{12} & f_{12}^{2}-g_{11}w_{22} & f_{12}f_{13}-g_{11}w_{23}
	\end{array}\right).
\end{equation}
Since 
\[
f_{12}^{2}-g_{11}w_{22}\neq0,
\]
$W_{i_{2}j}^{(2)}$ has no non-trivial null-eigenvector. The algorithm
ends.

In this case, we have three constraints $\phi_{1}^{(0)},\phi_{2}^{(0)}$
and $\phi^{(1)}$ given in (\ref{Lcons_1d2v_c3_phi1}), (\ref{Lcons_1d2v_c3_phi2})
and (\ref{Lcons_1d2a_c3.1_phi1}), one gauge identity $G^{(1)}$ given
in (\ref{Lcons_1d2a_c3.1_G1}), respectively. Therefore there is no
dynamical degree of freedom.

\subsection{Case 3.2: level-2 \label{subsec:Lcons_1d2v_L2_c3.2}}

From (\ref{Lcons_1d2v_c3.2_phi11}) and (\ref{Lcons_1d2v_c3.2_phi12}),
by appending $\dot{\phi}^{(1)}$ and $\dot{\phi}^{(2)}$ to $\mathcal{E}_{i_{1}}^{(1)}$
we get
\begin{equation}
	\mathcal{E}_{i_{2}}^{(2)}=W_{i_{2}j}^{(2)}\ddot{q}^{j}+V_{i_{2}}^{(2)},
\end{equation}
with
\begin{equation}
	W_{i_{2}j}^{(2)}=\left(\begin{array}{ccc}
		g_{11} & 0 & 0\\
		0 & 0 & 0\\
		0 & 0 & 0\\
		-f_{12} & 0 & 0\\
		-f_{13} & 0 & 0\\
		-g_{11}w_{12} & f_{12}^{2}-g_{11}w_{22} & f_{12}f_{13}-g_{11}w_{23}\\
		-g_{11}w_{13} & f_{13}f_{12}-g_{11}w_{23} & f_{13}^{2}-g_{11}w_{33}
	\end{array}\right),
\end{equation}
and
\begin{equation}
	V_{i_{2}}^{(2)}=\left(\begin{array}{c}
		f_{12}\dot{q}^{2}+f_{13}\dot{q}^{3}-w_{11}q^{1}-w_{12}q^{2}-w_{13}q^{3}\\
		-f_{12}\dot{q}^{1}-w_{12}q^{1}-w_{22}q^{2}-w_{23}q^{3}\\
		-f_{13}\dot{q}^{1}-w_{13}q^{1}-w_{23}q^{2}-w_{33}q^{3}\\
		-w_{12}\dot{q}^{1}-w_{22}\dot{q}^{2}-w_{23}\dot{q}^{3}\\
		-w_{13}\dot{q}^{1}-w_{23}\dot{q}^{2}-w_{33}\dot{q}^{3}\\
		-f_{12}w_{11}\dot{q}^{1}-f_{12}w_{12}\dot{q}^{2}-f_{12}w_{13}\dot{q}^{3}\\
		-f_{13}w_{11}\dot{q}^{1}-f_{13}w_{12}\dot{q}^{2}-f_{13}w_{13}\dot{q}^{3}
	\end{array}\right).
\end{equation}
We have to examine whether $W_{i_{2}j}^{(2)}$ possesses new non-trivial
null-eigenvectors. According to the rank of the matrix $\bm{\Delta}$
in (\ref{Lcons_L1_c3_Delta}), we have to discuss three sub-cases.

\subsubsection{Case 3.2.1: $\det\Delta\protect\neq0$}

In this case, clearly there is no non-trivial eigenvectors of $W_{i_{2}j}^{(2)}$.
The algorithm ends.

\subsubsection{Case 3.2.2: $\mathrm{rank}\Delta=1$ and $\mathcal{D}^{(1)}\protect\neq0$}

In this case $\det\bm{\Delta}=0$ but $\bm{\Delta}\neq0$. Similar
to the discussion in Sec. \ref{par:Lcons_1d1a_L1_c3.1}, we make use
of the form (\ref{Lcons_1d2a_L1_c3.1_gDform}) and keep in mind that
$\omega\neq0$. Then $W_{i_{2}j}^{(2)}$ redueces to 
\begin{equation}
	W_{i_{2}j}^{(2)}\rightarrow\left(\begin{array}{ccc}
		g_{11} & 0 & 0\\
		0 & 0 & 0\\
		0 & 0 & 0\\
		-f_{12} & 0 & 0\\
		-f_{13} & 0 & 0\\
		-g_{11}w_{12} & \omega & \omega\,\eta\\
		-g_{11}w_{13} & \omega\,\eta & \omega\,\eta^{2}
	\end{array}\right).
\end{equation}
$W_{i_{2}j}^{(2)}$ possesses a single non-trivial null-eigenvector,
which can be chosen to be
\begin{equation}
	u_{i_{2}}^{(2)}:=\left(\begin{array}{c}
		-\eta\,w_{12}+w_{13}\\
		0\\
		0\\
		0\\
		0\\
		-\eta\\
		1
	\end{array}\right).\label{Lcons_L2_c3.2.2_u2}
\end{equation}

Contracting $u_{i_{2}}^{(2)}$ with $\mathcal{E}_{i_{2}}^{(2)}$ yields
\begin{eqnarray}
	u^{(2)i_{2}}\mathcal{E}_{i_{2}}^{(2)} & = & \left(\eta\,f_{12}w_{11}-f_{13}w_{11}\right)\dot{q}^{1}+\left(f_{12}w_{13}-f_{13}w_{12}\right)\dot{q}^{2}+\eta\left(f_{12}w_{13}-f_{13}w_{12}\right)\dot{q}^{3}\nonumber \\
	&  & +w_{11}\left(\eta\,w_{12}-w_{13}\right)q^{1}+w_{12}\left(\eta\,w_{12}-w_{13}\right)q^{2}+w_{13}\left(\eta\,w_{12}-w_{13}\right)q^{3}.
\end{eqnarray}
Comparing with $\phi_{1}^{(0)},\phi_{2}^{(0)}$ in (\ref{Lcons_1d2v_c3_phi1})-(\ref{Lcons_1d2v_c3_phi2})
and $\phi_{1}^{(1)},\phi_{2}^{(1)}$ in (\ref{Lcons_1d2v_c3.2_phi11})-(\ref{Lcons_1d2v_c3.2_phi12}),
after some manipulations, one find that in this case $u^{(2)i_{2}}\mathcal{E}_{i_{2}}^{(2)}$
always leads to a new constraint at ``level-2'':
\begin{equation}
	\phi^{(2)}:=u^{(2)i_{2}}\mathcal{E}_{i_{2}}^{(2)}\approx0.\label{Lcons_1d2a_L2_c3.2.2_phi2}
\end{equation}

\subsubsection{Case 3.2.3: $\Delta=0$}

In this case clearly there are two new non-trivial null-eigenvectors
for $W_{i_{2}j}^{(2)}$:
\begin{equation}
	u_{1,i_{2}}^{(2)}=\left(\begin{array}{c}
		w_{12}\\
		0\\
		0\\
		0\\
		0\\
		1\\
		0
	\end{array}\right),\qquad u_{2,i_{2}}^{(2)}=\left(\begin{array}{c}
		w_{13}\\
		0\\
		0\\
		0\\
		0\\
		0\\
		1
	\end{array}\right).
\end{equation}
Contracting $u_{1,i_{2}}^{(2)}$ and $u_{2,i_{2}}^{(2)}$ with $\mathcal{E}_{i_{2}}^{(2)}$
yields
\begin{equation}
	u_{1}^{(2)i_{2}}\mathcal{E}_{i_{2}}^{(2)}=-f_{12}w_{11}\dot{q}^{1}+\left(w_{12}f_{13}-f_{12}w_{13}\right)\dot{q}^{3}-w_{12}w_{11}q^{1}-w_{12}^{2}q^{2}-w_{12}w_{13}q^{3},
\end{equation}
and
\begin{eqnarray}
	u_{2}^{(2)i_{2}}\mathcal{E}_{i_{2}}^{(2)} & = & -f_{13}w_{11}\dot{q}^{1}+\left(w_{13}f_{12}-f_{13}w_{12}\right)\dot{q}^{2}-w_{13}w_{11}q^{1}-w_{13}w_{12}q^{2}-w_{13}^{2}q^{3}.
\end{eqnarray}
Since we have already assumed $\bm{\Delta}=0$, (\ref{Lcons_1d2v_2id_fw})
must be satisfied, i.e., $w_{12}f_{13}-f_{12}w_{13}\neq0$.

Comparing with the constraints $\phi_{1}^{(0)},\phi_{2}^{(0)}$ in
(\ref{Lcons_1d2v_c3_phi1}) and (\ref{Lcons_1d2v_c3_phi2}), $\phi_{1}^{(1)},\phi_{2}^{(1)}$
in (\ref{Lcons_1d2v_c3.2_phi11}) and (\ref{Lcons_1d2v_c3.2_phi12}),
since the determinant of the $6\times6$ matrix (after making use
of (\ref{Lcons_1d2v_2id_w22w23w33}))
\begin{equation}
	\left(\begin{array}{cccccc}
		-f_{12} & 0 & 0 & -w_{12} & -\frac{f_{12}^{2}}{g_{11}} & -\frac{f_{12}f_{13}}{g_{11}}\\
		-f_{13} & 0 & 0 & -w_{13} & -\frac{f_{12}f_{13}}{g_{11}} & -\frac{f_{13}^{2}}{g_{11}}\\
		-g_{11}w_{12} & 0 & 0 & -f_{12}w_{11} & -f_{12}w_{12} & -f_{12}w_{13}\\
		-g_{11}w_{13} & 0 & 0 & -f_{13}w_{11} & -f_{13}w_{12} & -f_{13}w_{13}\\
		-f_{12}w_{11} & 0 & w_{12}f_{13}-f_{12}w_{13} & -w_{12}w_{11} & -w_{12}^{2} & -w_{12}w_{13}\\
		-f_{13}w_{11} & w_{13}f_{12}-f_{13}w_{12} & 0 & -w_{13}w_{11} & -w_{13}w_{12} & -w_{13}^{2}
	\end{array}\right)
\end{equation}
is
\begin{equation}
	-\left(w_{12}f_{13}-f_{12}w_{13}\right)^{5}\neq0,
\end{equation}
$u_{1}^{(2)i_{2}}\mathcal{E}_{i_{2}}^{(2)}$ and $u_{2}^{(2)i_{2}}\mathcal{E}_{i_{2}}^{(2)}$
lead to two new constraints at ``level-2'':
\begin{eqnarray}
	\phi_{1}^{(2)}:=u_{1}^{(2)i_{2}}\mathcal{E}_{i_{2}}^{(2)} & \approx & 0,\label{Lcons_1d2v_c3.2.3_phi1}\\
	\phi_{2}^{(2)}:=u_{2}^{(2)i_{2}}\mathcal{E}_{i_{2}}^{(2)} & \approx & 0.\label{Lcons_1d2v_c3.2.3_phi2}
\end{eqnarray}

\subsection{Case 3.2.2: level-3 \label{subsec:Lcons_1d2v_L3_c3.2.2}}

From (\ref{Lcons_1d2a_L2_c3.2.2_phi2}), we have
\begin{equation}
	\mathcal{E}_{i_{3}}^{(3)}=W_{i_{3}j}^{(3)}\ddot{q}^{j}+V_{i_{3}}^{(3)},
\end{equation}
with
\begin{equation}
	W_{i_{3}j}^{(3)}=\left(\begin{array}{ccc}
		g_{11} & 0 & 0\\
		0 & 0 & 0\\
		0 & 0 & 0\\
		-f_{12} & 0 & 0\\
		-f_{13} & 0 & 0\\
		-g_{11}w_{12} & \omega & \omega\,\eta\\
		-g_{11}w_{13} & \omega\,\eta & \omega\,\eta^{2}\\
		\eta f_{12}w_{11}-f_{13}w_{11} & f_{12}w_{13}-f_{13}w_{12} & \eta\left(f_{12}w_{13}-f_{13}w_{12}\right)
	\end{array}\right),
\end{equation}
and
\begin{equation}
	V_{i_{3}j}^{(3)}=\left(\begin{array}{c}
		f_{12}\dot{q}^{2}+f_{13}\dot{q}^{3}-w_{11}q^{1}-w_{12}q^{2}-w_{13}q^{3}\\
		\frac{q^{2}\left(\omega-f_{12}^{2}\right)}{g_{11}}+\frac{q^{3}(\eta\omega-f_{12}f_{13})}{g_{11}}-f_{12}\dot{q}^{1}-w_{12}q^{1}\\
		\frac{q^{2}(\eta\omega-f_{12}f_{13})}{g_{11}}+\frac{q^{3}\left(\eta^{2}\omega-f_{13}^{2}\right)}{g_{11}}-f_{13}\dot{q}^{1}-w_{13}q^{1}\\
		\frac{\dot{q}^{2}\left(\omega-f_{12}^{2}\right)}{g_{11}}+\frac{\dot{q}^{3}(\eta\omega-f_{12}f_{13})}{g_{11}}-w_{12}\dot{q}^{1}\\
		\frac{\dot{q}^{2}(\eta\omega-f_{12}f_{13})}{g_{11}}+\frac{\dot{q}^{3}\left(\eta^{2}\omega-f_{13}^{2}\right)}{g_{11}}-w_{13}\dot{q}^{1}\\
		-f_{12}w_{11}\dot{q}^{1}-f_{12}w_{12}\dot{q}^{2}-f_{12}w_{13}\dot{q}^{3}\\
		-f_{13}w_{11}\dot{q}^{1}-f_{13}w_{12}\dot{q}^{2}-f_{13}w_{13}\dot{q}^{3}\\
		w_{11}\dot{q}^{1}(\eta w_{12}-w_{13})+w_{12}\dot{q}^{2}(\eta w_{12}-w_{13})-w_{13}\dot{q}^{3}(w_{13}-\eta w_{12})
	\end{array}\right).
\end{equation}

There is one non-trivial null-eigenvector for $W_{i_{3}j}^{(3)}$,
which can be chosen to be
\begin{equation}
	u_{i_{3}}^{(3)}=\left(\begin{array}{c}
		0\\
		0\\
		0\\
		\frac{g_{11}w_{12}}{f_{12}}\left(f_{12}w_{13}-f_{13}w_{12}\right)+w_{11}\omega\left(\eta-\frac{f_{13}}{f_{12}}\right)\\
		0\\
		f_{13}w_{12}-f_{12}w_{13}\\
		0\\
		\omega
	\end{array}\right)\label{Lcons_1d2a_L3_c3.2.2_u3a}
\end{equation}
for $f_{12}\neq0$, and to be
\begin{equation}
	u_{i_{3}}^{(3)}=\left(\begin{array}{c}
		0\\
		0\\
		0\\
		0\\
		g_{11}\frac{w_{12}}{f_{13}}\left(f_{12}w_{13}-f_{13}w_{12}\right)+\omega\frac{w_{11}}{f_{13}}\left(\eta f_{12}-f_{13}\right)\\
		f_{13}w_{12}-f_{12}w_{13}\\
		0\\
		\omega
	\end{array}\right)\label{Lcons_1d2a_L3_c3.2.2_u3b}
\end{equation}
for $f_{13}\neq0$, and to be
\begin{equation}
	u_{i_{3}}^{(3)}=\left(\begin{array}{c}
		0\\
		0\\
		0\\
		0\\
		0\\
		0\\
		0\\
		1
	\end{array}\right),\label{Lcons_1d2a_L3_c3.2.2_u3c}
\end{equation}
for $f_{12}=f_{13}=0$. In all cases, the contraction of $u_{i_{3}}^{(3)}$
with $\mathcal{E}_{i_{3}}^{(3)}$ can be shown to be a new constraint
at ``level-3''
\begin{equation}
	\phi^{(3)}:=u^{(3)i_{3}}\mathcal{E}_{i_{3}}^{(3)}\approx 0,
\end{equation}
since the $6\times6$ matrix $\bm{M}^{(3)}$ defined by
\begin{equation}
	\left(\begin{array}{c}
		\phi_{1}^{(0)}\\
		\phi_{2}^{(0)}\\
		\phi_{1}^{(1)}\\
		\phi_{2}^{(1)}\\
		\phi^{(2)}\\
		u^{(3)i_{3}}\mathcal{E}_{i_{3}}^{(3)}
	\end{array}\right)=\bm{M}^{(3)}\left(\begin{array}{c}
		\dot{q}^{1}\\
		\dot{q}^{2}\\
		\dot{q}^{3}\\
		q^{1}\\
		q^{2}\\
		q^{3}
	\end{array}\right),
\end{equation}
always possesses a non-vanishing determinant. 

\subsection{Case 3.2.2: level-4 \label{subsec:Lcons_1d2v_L4_c3.2.2}}

There is no non-trivial null-eigenvector of $W_{i_{4}j}^{(4)}$. The
algorithm ends.

To summarize, we have six constraints: $\phi_{1}^{(0)},\phi_{2}^{(0)},\phi_{1}^{(1)},\phi_{2}^{(1)},\phi^{(2)},\phi^{(3)}$,
and thus there is no dynamical degree of freedom.

\subsection{Case 3.2.3: level-3 \label{subsec:Lcons_1d2v_L3_c3.2.3}}

From (\ref{Lcons_1d2v_c3.2.3_phi1}) and (\ref{Lcons_1d2v_c3.2.3_phi2}),
by appending $\dot{\phi}_{1}^{(2)}$ and $\dot{\phi}_{2}^{(2)}$ to
$\mathcal{E}_{i_{2}}^{(2)}$, we get
\begin{equation}
	\mathcal{E}_{i_{3}}^{(3)}=W_{i_{3}j}^{(3)}\ddot{q}^{j}+V_{i_{3}}^{(3)},
\end{equation}
with (we have used (\ref{Lcons_1d2v_2id_w22w23w33}) to replace $w_{22},w_{23}$
and $w_{33}$)
\begin{equation}
	W_{i_{3}j}^{(3)}=\left(\begin{array}{ccc}
		g_{11} & 0 & 0\\
		0 & 0 & 0\\
		0 & 0 & 0\\
		-f_{12} & 0 & 0\\
		-f_{13} & 0 & 0\\
		-g_{11}w_{12} & 0 & 0\\
		-g_{11}w_{13} & 0 & 0\\
		-f_{12}w_{11} & 0 & w_{12}f_{13}-f_{12}w_{13}\\
		-f_{13}w_{11} & w_{13}f_{12}-f_{13}w_{12} & 0
	\end{array}\right),
\end{equation}
and 
\begin{equation}
	V_{i_{2}}^{(2)}=\left(\begin{array}{c}
		f_{12}\dot{q}^{2}+f_{13}\dot{q}^{3}-w_{11}q^{1}-w_{12}q^{2}-w_{13}q^{3}\\
		-f_{12}\dot{q}^{1}-w_{12}q^{1}-\frac{f_{12}^{2}}{g_{11}}q^{2}-\frac{f_{12}f_{13}}{g_{11}}q^{3}\\
		-f_{13}\dot{q}^{1}-w_{13}q^{1}-\frac{f_{12}f_{13}}{g_{11}}q^{2}-\frac{f_{13}^{2}}{g_{11}}q^{3}\\
		-w_{12}\dot{q}^{1}-\frac{f_{12}^{2}}{g_{11}}\dot{q}^{2}-\frac{f_{12}f_{13}}{g_{11}}\dot{q}^{3}\\
		-w_{13}\dot{q}^{1}-\frac{f_{12}f_{13}}{g_{11}}\dot{q}^{2}-\frac{f_{13}^{2}}{g_{11}}\dot{q}^{3}\\
		-f_{12}w_{11}\dot{q}^{1}-f_{12}w_{12}\dot{q}^{2}-f_{12}w_{13}\dot{q}^{3}\\
		-f_{13}w_{11}\dot{q}^{1}-f_{13}w_{12}\dot{q}^{2}-f_{13}w_{13}\dot{q}^{3}\\
		-w_{12}w_{11}\dot{q}^{1}-w_{12}^{2}\dot{q}^{2}-w_{12}w_{13}\dot{q}^{3}\\
		-w_{13}w_{11}\dot{q}^{1}-w_{13}w_{12}\dot{q}^{2}-w_{13}^{2}\dot{q}^{3}
	\end{array}\right).
\end{equation}
Since $w_{12}f_{13}-f_{12}w_{13}\neq0$, there is no non-trivial eigenvector
for $W_{i_{3}j}^{(3)}$. The algorithm ends.

To summarize, in this case, we have six constraints: $\phi_{1}^{(0)},\phi_{2}^{(0)}$
in (\ref{Lcons_1d2v_c3_phi1}) and (\ref{Lcons_1d2v_c3_phi2}), $\phi_{1}^{(1)},\phi_{2}^{(1)}$
in (\ref{Lcons_1d2v_c3.2_phi11}) and (\ref{Lcons_1d2v_c3.2_phi12}),
$\phi_{1}^{(2)},\phi_{2}^{(2)}$ in (\ref{Lcons_1d2v_c3.2.3_phi1})-(\ref{Lcons_1d2v_c3.2.3_phi2}),
respectively. Therefore there is no dynamical degree of freedom.

\subsection{Summary}

The classification of Lagrangians with one dynamical and two auxiliary variables is summarized in Tab. \ref{tab:Lcons_1d2a_cls}. According to the types of constraints/identities as well as to that at which level these constraints/identities arise, there are in total 8 cases (in
some sense 8 types of theories). 
Counting the number of degrees of freedom in the Lagrangian approach is discussed in 
\cite{Pons:1986zg,Gracia:1988xp,Kim:1998wm,Henneaux:1990au}, which is given by the simple formula \cite{Diaz:2014yua}
	\begin{equation}
		\#_{\mathrm{D.O.F.}} = N-\frac{1}{2}(l+g+e),
	\end{equation}
in which $N$, $l$ and $g$ are the total numbers of the variables, the Lagrangian constraints and the gauge identities, respectively. $e$ is the total number of the gauge parameters plus its successive derivatives.
	\begin{table}[H]
		\begin{centering}
			\begin{tabular}{|l|c|c|c|c|c|}
				\hline 
				& level-0 & level-1 & level-2 & level-3 & $\#_{\mathrm{D.O.F.}}$\tabularnewline
				\hline 
				case 1 & $G_{1}^{(0)},G_{2}^{(0)}$ & - & - & - & 1\tabularnewline
				\hline 
				case 2.1 & \multirow{3}{*}{$\phi^{(0)},G^{(0)}$} & $G^{(1)}$ & - & - & 0\tabularnewline
				\cline{1-1} \cline{3-6} \cline{4-6} \cline{5-6} \cline{6-6} 
				case 2.2.1 &  & \multirow{2}{*}{$\phi^{(1)}$} & - & - & 1\tabularnewline
				\cline{1-1} \cline{4-6} \cline{5-6} \cline{6-6} 
				case 2.2.2 &  &  & $\phi^{(2)}$ & $\phi^{(3)}$ & 0\tabularnewline
				\hline 
				case 3.1 & \multirow{4}{*}{$\phi_{1}^{(0)},\phi_{2}^{(0)}$} & $\phi^{(1)},G^{(1)}$ & - & - & 0\tabularnewline
				\cline{1-1} \cline{3-6} \cline{4-6} \cline{5-6} \cline{6-6} 
				case 3.2.1 &  & \multirow{3}{*}{$\phi_{1}^{(1)},\phi_{2}^{(1)}$} & - & - & 1\tabularnewline
				\cline{1-1} \cline{4-6} \cline{5-6} \cline{6-6} 
				case 3.2.2 &  &  & $\phi^{(2)}$ & $\phi^{(3)}$ & 0\tabularnewline
				\cline{1-1} \cline{4-6} \cline{5-6} \cline{6-6} 
				case 3.2.3 &  &  & $\phi_{1}^{(2)},\phi_{2}^{(2)}$ & - & 0\tabularnewline
				\hline 
			\end{tabular}
			\par\end{centering}
		\caption{Classification of the quadratic Lagrangians with one dynamical and two auxiliary
			variables.}
		\label{tab:Lcons_1d2a_cls}
	\end{table}
With this classification, it is transparent that the linear scalar perturbations in GR belong to ``case 3.1''.
While what we explored in this work corresponds to ``case 3.2.2'' and ``case 3.2.3'' (together with ``case 3.1'').

%

\end{document}